\documentclass[10pt,a4paper]{article}

\usepackage[english]{babel}
\usepackage{a4wide}
\usepackage{amssymb,amsmath,feynmf,indentfirst}
\usepackage{pgf, subfig,booktabs}

\newcommand{\pslash}{ \mkern-6mu \not \mkern-4mu p }
\newcommand{\mpi}{m_\pi}
\newcommand{\po}{p_1}
\newcommand{\pt}{p_2}
\newcommand{\pth}{p_3}

\newcommand{\fpi}{f_\pi}
\newcommand{\ga}{g_A}
\newcommand{\kn}{\mathcal{K}}
\newcommand{\otp}{\po + \pt}
\newcommand{\otm}{\po - \pt}
\newcommand{\tom}{\pt - \po}
\newcommand{\othp}{\po + \pth}
\newcommand{\othm}{\po - \pth}
\newcommand{\tthp}{\pt + \pth}
\newcommand{\tthm}{\pt - \pth}

\title{Chiral thermodynamics of nuclear matter\thanks{Work supported in part by BMBF, GSI and the DFG Cluster of Excellence ``Origin and Structure of the Universe".}}

\author{Salvatore Fiorilla, Norbert Kaiser, Wolfram Weise \\
\small{Physik-Department, Technische Universit\"{a}t M\"{u}nchen, D-85747 Garching, Germany}}

\begin{document}

\maketitle

\begin{abstract}
The free energy and the equation of state of isospin-asymmetric nuclear matter are calculated at finite temperature up to three loop order in the framework of in-medium chiral perturbation theory, systematically incorporating one- and two-pion exchange dynamics to this order. Effects from the $ 2\pi$-exchange with explicit $ \Delta$-isobar excitation are included, as well as three-body forces. We construct the phase diagram of nuclear matter for different proton fractions $ x_p $ and investigate the dependence of nuclear matter properties on the isospin-asymmetry. A detailed study of the liquid-gas phase transition is performed. For isospin-symmetric nuclear matter we find a critical temperature of 15.1 MeV; as the isospin-asymmetry is increased, the liquid-gas coexistence region decreases until it disappears at $ x_p \simeq 0.05 $, while nuclear matter becomes unbound at $ x_p \simeq 0.12 $. The quadratic expansion of the free energy in the asymmetry parameter $ \delta $ is a good approximation at low temperature even for large $ \delta $. An estimate of chiral four-body correlations in nuclear and neutron matter is also performed and the corrections from such four-body interactions are found to be small. \\ \\
Keywords: chiral perturbation theory; nuclear matter; equation of state; thermodynamics; phase diagram; liquid-gas phase transition.
\end{abstract}

\section{Introduction and framework}

Theories and models that describe the thermodynamic properties of nuclear matter play an important role in applications to heavy-ion collisions and astrophysics. The plateau exhibited by the caloric curve of the nuclear fragments in nucleus-nucleus collisions is interpreted as the trace of a first-order liquid-gas phase transition \cite{gross,pocho,natowitz2}. Concerning astrophysics, the recent observation of a two-solar-mass neutron star \cite{demorest} provides important constraints for the equation of state of strongly interacting matter, as it would rule out many exotic models which tend to produce too soft equations of state. 

Within the last two decades a novel approach to the nuclear many-body problem has been developed, based on the understanding of the nucleon-nucleon interaction in the framework of chiral effective field theories \cite{lutz, kaiser1, kaiser2, kaiser3}. This approach relies on the scale separation between the short-range physics and the long- and intermediate-distance dynamics, the latter responsible for the attractive part of the nucleon-nucleon interaction described  primarily by $ 2\pi$-exchange dynamics. While the long- and intermediate-range, pion-driven dynamics (in the presence of the nuclear medium) are treated explicitly, the unresolved short-distance physics  is encoded in a few contact terms with coefficients fixed in order to reproduce selected known bulk properties of nuclear matter. Many other ground state and single-particle properties and nuclear thermodynamics emerge as predictions.

In the following we briefly recall the theoretical framework of in-medium Chiral Perturbation Theory (ChPT) developed so far for isospin-symmetric nuclear matter. In the present work we generalize such calculations to isospin-asymmetric nuclear matter with an arbitrary proton-to-neutron ratio and perform a systematic analysis of the thermodynamic properties of the resulting equation of state. This short summary of the basic framework will be restricted to the essentials since the more detailed discussion can be found in Refs.~\cite{kaiser1, kaiser2, kaiser3}.

The relevant active degrees of freedom for the description of the nuclear many-body problem are nucleons and pions. In-medium ChPT is applicable to nuclear many-body systems as long as the Fermi momentum, $ k_F $, is small compared to the scale of spontaneous chiral symmetry breaking in QCD, $ \Lambda_\chi \sim 4\pi f_\pi \sim 1 $ GeV, where $ f_\pi $ is the pion decay constant in vacuum. The characteristic range of nuclear momenta is comparable to about twice the pion mass ($ k_{F0} \approx 2\,m_\pi $, where $ k_{F0} \simeq 263 $ MeV is the Fermi momentum at the saturation density of nuclear matter, $ \rho_0 \simeq 0.16\,\text{fm}^{-3} $). This implies that $ 1\pi$- and $ 2\pi$-exchange processes have to be treated explicitly in the presence of the nuclear medium. Moreover, to  complete the picture it is mandatory to include the $ \Delta(1232)$-isobar excitation as an explicit degree of freedom, because the splitting between the $ \Delta$-isobar mass and the nucleon mass, $ \Delta = 293 $ MeV, is again comparable to $ 2 \,m_\pi $. Consequently, the ``small scales''  that appear in our calculation scheme are  the Fermi momentum of the nucleons, $ k_F $, the pion mass $ m_\pi $,  and the $ \Delta(1232)-N $ mass splitting. The energy per particle at $ T = 0 $ is then derived as a systematic expansion in powers of $ k_F $. The expansion coefficients are non-trivial functions of the ratios $ k_F / m_\pi $ and of $ \Delta/m_\pi $. 

Explicit $ 1\pi $- and $ 2\pi $-exchange dynamics in the medium are so far treated up to three-loop order in the energy density \cite{kaiser1,kaiser2,kaiser3}. The basic ingredient in performing calculations at finite baryon density is the in-medium nucleon propagator. Consider first the case of zero temperature. Instead of the usual decomposition of the propagator into a particle and a hole part, it is convenient to use the following equivalent representation:
\begin{equation} \label{propagator}
S_N(p) = \left(\, \pslash + M_N \right) \left[ \frac{i}{p^2- M_N^2 + i\epsilon} - 2 \pi \delta (p^2-M_N^2) \theta (k_F - |\mathbf{p}|) \theta(p_0) \right]   \ ,
\end{equation}
where $ p^\mu = (p_0, \mathbf{p}) $ and $ M_N = 939 $ MeV is (free) the nucleon mass.  The first term is the vacuum propagator of the nucleon, while the second term, the medium insertion, takes into account the presence of a filled Fermi sea of nucleons.
The calculation is then organized according to the number of medium insertions in the diagrams representing the energy density. Diagrams with no medium insertion give an unobservable shift of the vacuum energy. Diagrams with one medium insertion renormalize the nucleon mass to its physical value and provide a description at the level of a (non-interacting) Fermi gas. Diagrams with two and three medium insertions produce the interesting many-body effects. Two-body terms involving the nucleon-nucleon $ T$-matrix come from diagrams with two medium insertions. Three medium insertions generate three-body terms such as  Pauli-blocking effects on the two-body terms, and so forth. Convergence in the series of medium insertions, or powers of $ k_F $, at the three-loop level is realized for $ k_F \ll \Lambda_\chi \sim 1 $ GeV as long as four-nucleon correlations are not prominent. Effects of four-body correlations are still an open issue to be explored in further (ongoing) studies \cite{epelbaum}. Genuine chiral four-body contributions to the energy per particle are examined later in Section \ref{sec5}. They turn out to be negligibly small.

 \begin{figure}[tbp]
\centering
 \subfloat[][]{\label{1pion_1}
\includegraphics[height=2.4cm]{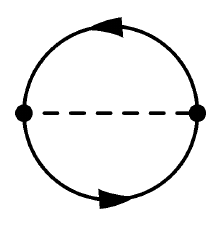}} 
 \subfloat[][]{\label{1pion_2}
 \includegraphics[height=2.4cm]{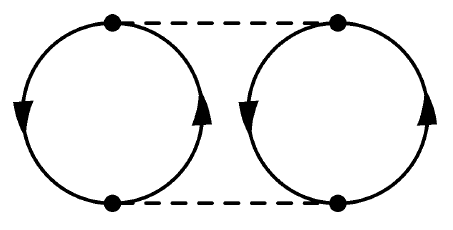}} 
  \subfloat[][]{\label{1pion_3}
   \includegraphics[height=2.4cm]{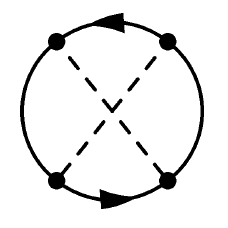}} 
\caption{ (a): $ 1\pi$-exchange Fock diagram; (b): iterated $ 1\pi$-exchange Hartree diagram; (c): iterated $ 1\pi$-exchange Fock diagram. Diagram (a) includes two medium insertions, diagrams (b) and (c) involve two and three medium insertions.}
\label{1pion}
\end{figure}

 \begin{figure}[tbp]
\centering
\includegraphics[height=2.6cm]{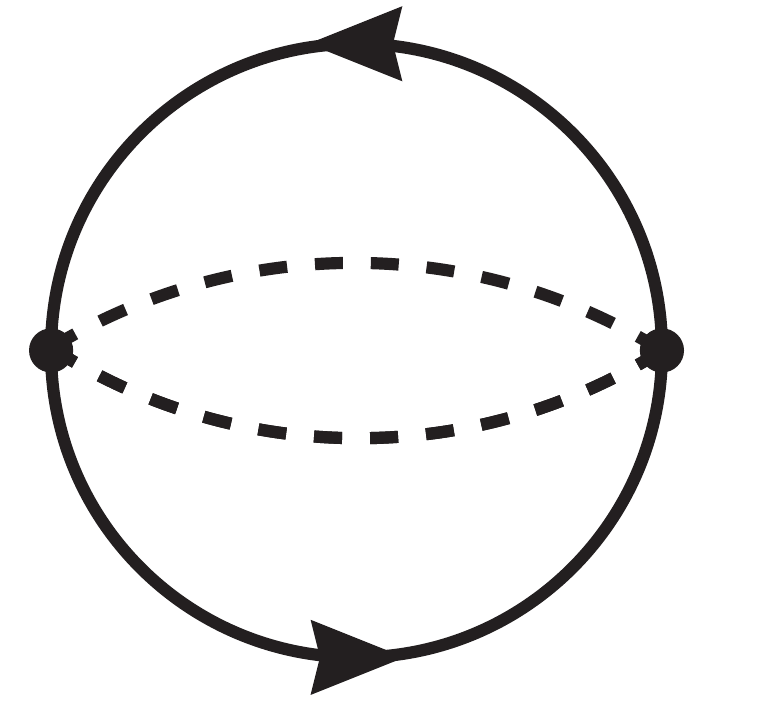}
\includegraphics[height=2.6cm]{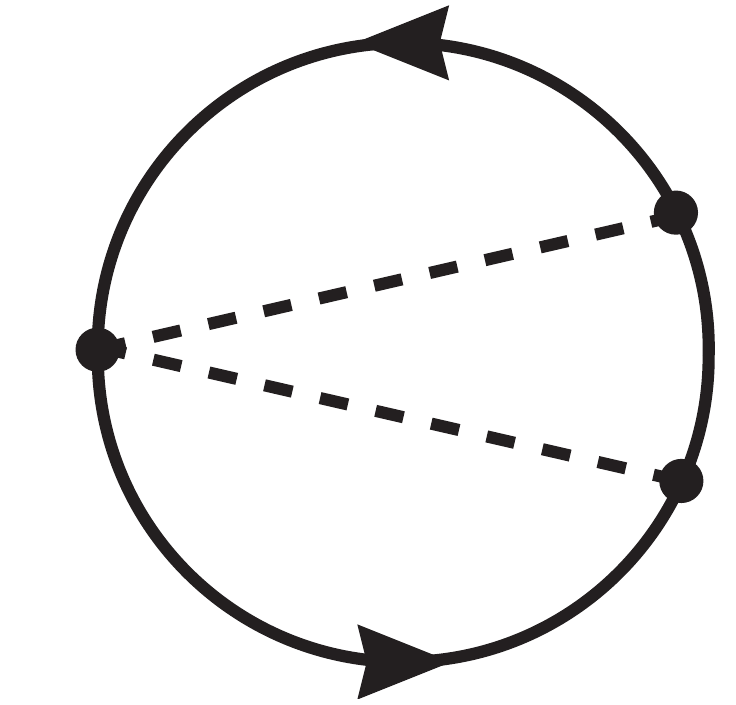}   
 \caption{Examples of irreducible $ 2\pi$-exchange diagrams to be evaluated with two medium insertions.}
\label{2pion}
\end{figure}

\begin{figure}[tbp] 
 \center
 \includegraphics[height=2.4cm]{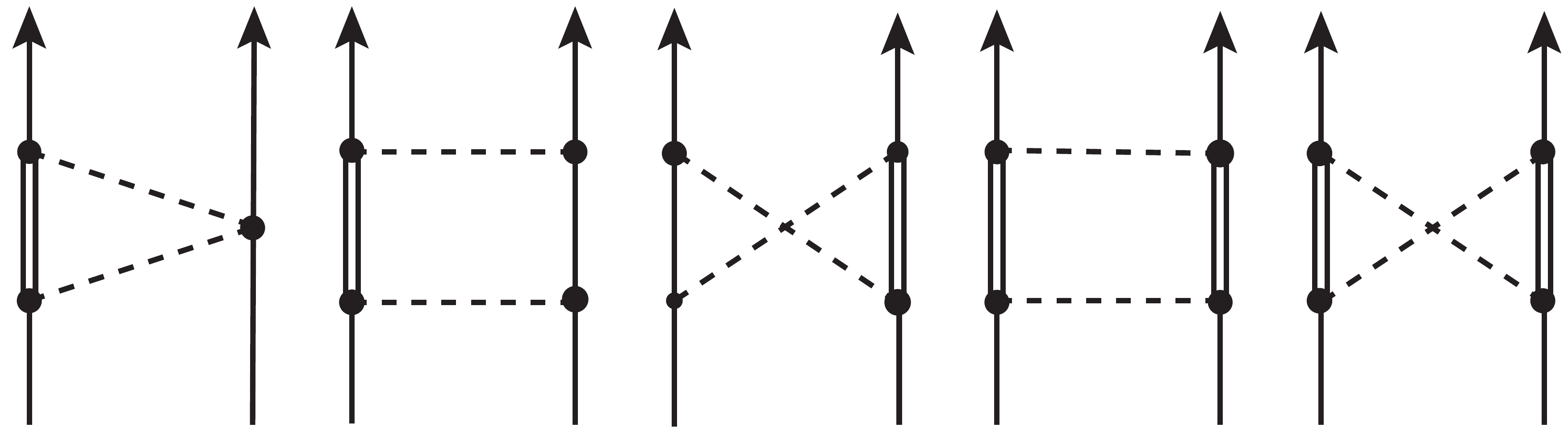}     
\caption{$ 2\pi$-exchange diagrams with single and double $ \Delta(1232)$-isobar excitation.}
\label{delta}
\end{figure}

\begin{figure}[tbp] 
 \center
 \includegraphics[height=2.4cm]{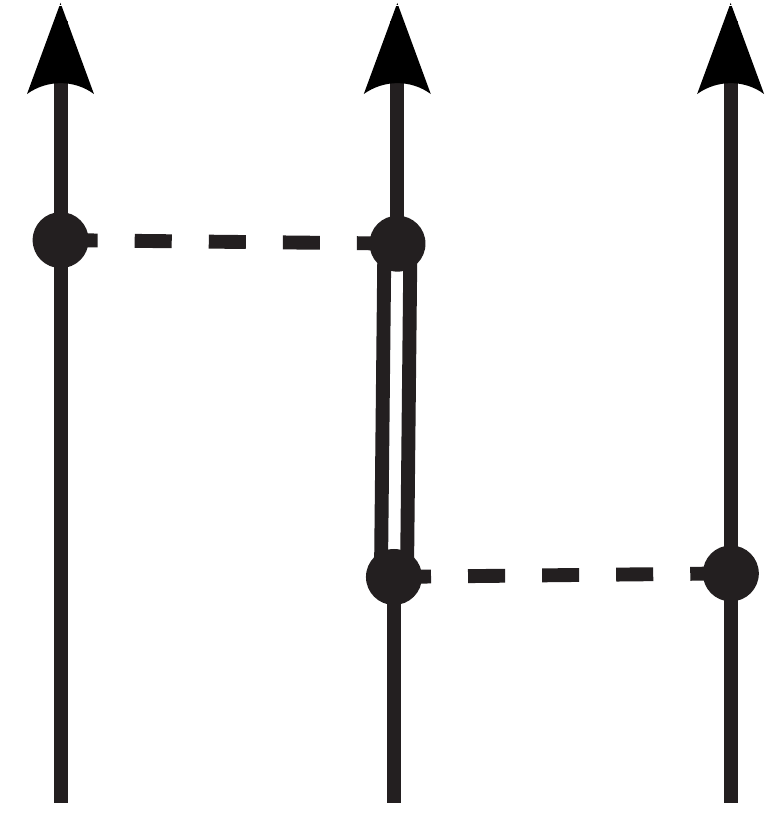}     
\caption{Three-nucleon term with intermediate $ \Delta(1232)$-excitation.}
\label{3nucleon}
\end{figure}

Fig.~\ref{1pion} shows the $ 1\pi$- and iterated $ 1\pi$-exchange diagrams evaluated with up to three medium insertions. The iterated one-pion exchange terms include the second-order tensor interaction that is well known to be very important in nuclear systems. Fig.~\ref{2pion} displays irreducible $ 2\pi$-exchange diagrams contributing to the energy density. The detailed discussion and terminology concerning iterated $ 1\pi$-exchange and irreducible $ 2\pi$-exchange diagrams in the context of  ChPT, both in vacuum and in-medium, can be found in Refs.~\cite{kaiser1, kaiser2, kaiser3}. The $ 2\pi$-exchange diagrams with inclusion of $ \Delta$-isobar excitations as intermediate states are shown in Fig.~\ref{delta}. Medium insertions on the second and third of these diagrams, cutting the intermediate nucleon line, lead to the three-nucleon term of Fig.~\ref{3nucleon}. 

Unresolved short-distance physics is encoded in two contact terms, a momentum independent term with coefficient $ B_3 $ and a $ p^2$-dependent term with coefficient $ B_5 $ \cite{kaiser3}. For a good description of nuclear matter properties an additional attractive three-nucleon contact interaction (controlled by the parameter $ \zeta $, see Eq.\eqref{zeta})  is required, which partly counteracts the  strongly repulsive $ \rho^2$-term in the energy per nucleon from the three-body diagram with $ \Delta$-isobar excitation. After tuning the minimum of the energy per particle of symmetric nuclear matter to $ -16 $ MeV, one can fix $ B_5 =  0 $.

The next step is the extension to finite temperatures. The ordering scheme in the number of medium insertions is now transcribed to the free energy density as a function of density $ \rho $ and temperature $ T $. The free energy density is written as the sum of convolution integrals:
\begin{multline} \label{convolution}
	\rho \, \bar{F}(\rho, T) = 4 \int\limits_0^\infty \mbox{d}p \, p \,\mathcal{K}_1(p) \, d(p) + \int\limits_0^\infty \mbox{d}p_1 \int\limits_0^\infty \mbox{d}p_2 \, \mathcal{K}_2(p_1,p_2) \, d(p_1) \, d(p_2) \\
	+ \int\limits_0^\infty \mbox{d}p_1 \int\limits_0^\infty \mbox{d}p_2 \int\limits_0^\infty \mbox{d}p_3 \, \mathcal{K}_3(p_1,p_2,p_3) \, d(p_1) \, d(p_2) \, d(p_3) + \rho \, \mathcal{\bar{A}}(\rho,T) \ ,
\end{multline}
where $  \bar{F}(\rho, T) $ is the free energy per nucleon and
\begin{equation}\label{density}
	d(p) = \frac{p}{2\,\pi^2} \left[ 1 + \exp{\frac{p^2 /2 M_N -\tilde{\mu}}{T}} \right]^{-1}
\end{equation}
is the density of nucleon states in momentum space. Complete expressions of the kernels $ \mathcal{K}_j $ can be found in \ref{appA}. The term $  \mathcal{\bar{A}}(\rho,T) $, the so-called anomalous contribution, is associated with a Fock term involving second order $ 1\pi$-exchange \cite{kaiser6}. This contribution vanishes at $ T = 0 $ and has its origin in the smoothing of the Fermi surface at $ T \neq 0 $. Its effect is small in the range of temperatures considered here. The one-body  effective ``chemical potential''  $ \tilde{\mu} $ in the distribution $ d(p)$ is related to the density through 
\begin{equation}
	\rho = 4 \int\limits_0^\infty \mbox{d}p \, p \, d(p) \ .
\end{equation}
The kernel $ \mathcal{K}_1 $ is  the contribution from the non-interacting nucleon gas, the kernels  $ \mathcal{K}_2 $ and $ \mathcal{K}_3 $ describe the effects of the interactions and include the sum of diagrams in Figs.~\ref{1pion}-\ref{delta} with two- and three-medium insertions.   The pressure is computed using the standard thermodynamic relation
\begin{equation} \label{pressure}
P(\rho, T) = \rho^2 \, \frac{\partial \bar{F}(\rho, T)}{\partial \rho} \ . 
\end{equation}

The calculation is then extended to asymmetric nuclear matter and neutron matter. In comparison with the calculation for symmetric nuclear matter, the only changes required are in isospin factors and the introduction of two new contact terms associated with the isospin dependence of the short-distance interactions, with coefficients denoted by $ B_{n,3} $ and $ B_{n,5} $; they are adjusted in order to reproduce the value of 34 MeV for the asymmetry energy at the saturation point \cite{kaiser3}.

 Note that for pure neutron matter the Pauli principle forbids a three-body contact interaction. The behaviour at low density is dictated by the  large neutron-neutron scattering length, $ a_{nn} \simeq 19 $ fm \cite{lenght1,lenght2}, and a resummation of the short-distance $ NN $ interaction to all orders in the presence of the medium is required for a realistic description at low densities ($ \rho_n \lesssim 0.02\,\text{fm}^{-3} $)\cite{kaiser4}. The full implementation of these effects is currently in progress.

In the present paper we extend the chiral three-loop calculation of isospin-symmetric nuclear matter to the isospin-asymmetric case and present a systematic investigation of the phase diagram of the corresponding equation of state for different values of the proton-to-neutron ratio. The Fermi seas of protons and neutrons are now filled unequally. The propagator \eqref{propagator} at $ T= 0 $ is split into proton and neutron contributions using isospin-projectors:
\begin{equation}
	\theta( k_F - |\mathbf{p}| ) \longrightarrow \frac{1+\tau_3}{2} \, \theta ( k_{p} - |\mathbf{p}| ) + \frac{1-\tau_3}{2} \, \theta (k_{n} - |\mathbf{p}| ) \ ,
\end{equation}
with $ k_{p} $ and $ k_{n} $ denoting, respectively, the Fermi momenta of protons and neutrons. The ``one-body'' chemical potentials of the two nucleon species in Eq.~\eqref{density} are now different.  At finite temperature we replace
\begin{equation}
d(p) \longrightarrow  \frac{1+\tau_3}{2} \, d_p(p) + \frac{1-\tau_3}{2} \, d_n(p) \ ,
\end{equation}
where $ d_p $ and $ d_n $ are the proton and neutron distributions. Each diagram now involves the sum of all possible combinations of proton and neutron medium insertions with their specific isospin factors. In the appendices we list the complete expressions for the diagrammatic expansion of the free energy density, both for isospin-symmetric (\ref{appA}) and asymmetric matter (\ref{appB}). \\

This paper is organized as follows. In section \ref{symmetric} we present and discuss the equation of  state of isospin-symmetric nuclear matter for different temperatures, featuring the liquid-gas phase transition. In section \ref{asymmetric} the calculations are extended to the isospin-asymmetric case, focusing on the behaviour of nuclear matter and its thermodynamic properties with varying proton-neutron asymmetry. In section \ref{asymmetry} we study the asymmetry free energy, its dependence on density and temperature  and the validity of the parabolic approximation for the free energy as a function of the asymmetry parameter, $ \delta = (\rho_n-\rho_p)/ (\rho_n+\rho_p) $. Finally, section \ref{summary} contains a summary and the conclusions.

\section{Equation of state of isospin-symmetric nuclear matter} \label{symmetric}

\begin{figure}[tbp] 
 \center
 \includegraphics[width=0.745\textwidth]{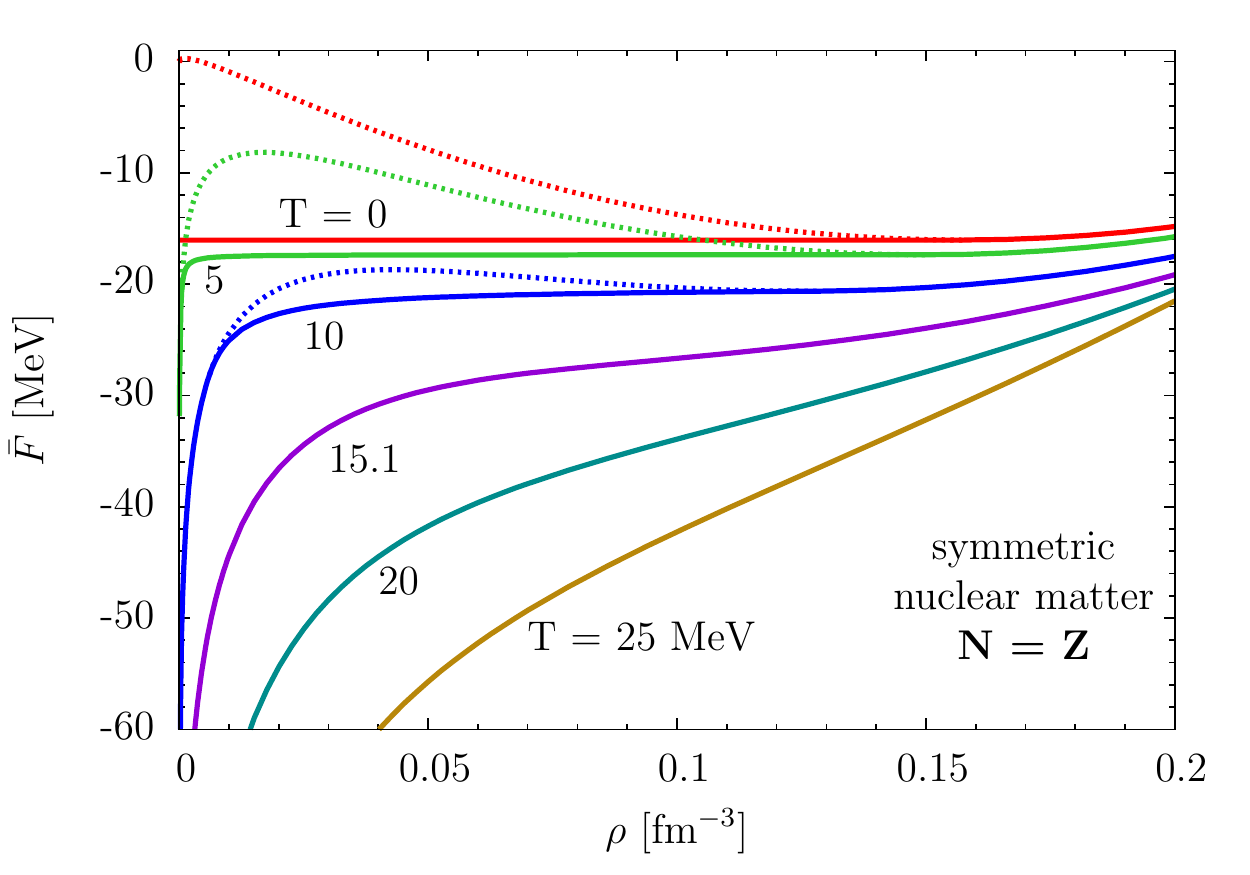}     
\caption{Free energy per particle of isospin-symmetric nuclear matter as a function of baryon density $ \rho $ for different temperatures. The dotted line indicates the non-physical behaviour of the free energy in the liquid-gas coexistence region. The physical free energy (solid lines) at low temperatures ($ T \lesssim 15 $ MeV) is obtained using the Maxwell construction.}
\label{fenergy_50}
\end{figure}

We start by recalling and extending the results found for isospin-symmetric nuclear matter in Ref.~\cite{kaiser3} using Eq.~\eqref{convolution}. The free energy per particle $ \bar{F} (\rho, T) $ as a function of density $ \rho $ for a sequence of temperatures up to 25 MeV is calculated using Eq.~\eqref{convolution} with the input for the interaction kernels $ \mathcal{K}_n $ specified in \ref{appA}. The result is shown in Fig.~\ref{fenergy_50}. The dotted lines indicate the non-physical behaviour of the equation of state in the liquid-gas first-order transition region. This part of each curve is substituted by the physical one (solid lines) using the Maxwell construction. 

At $ T = 0 $ the free energy equals the internal energy of the system. The minimum of the curve  (the saturation point) is located at $ \bar{E}_0 = -16.0 $ MeV, $ \rho_0 = 0.157\ \mbox{fm}^{-3} $, $ k_{F0} = 1.33\ \mbox{fm}^{-1} = 262 $ MeV. At finite temperatures the free energy displays a singular behaviour for $ \rho \rightarrow 0 $; this is a well-known generic feature that the present calculation shares with other types of many-body calculations \cite{pandha,horowitz}. 

The energy per particle of symmetric nuclear matter is usually expanded around the saturation point:  
\begin{equation} \label{par_eos}
	\bar{E}(\rho) \approx \bar{E}_0 + \frac{K}{2}  \left( \frac{\rho - \rho_0}{3 \rho_0} \right)^2 \ \ ,
\end{equation}
with $ K $ the compression modulus. From the curve at $ T = 0 $ we extract the value $ K \simeq 300 $ MeV, slightly larger than the values deduced from relativistic mean field models  \cite{vretenar} and from the systematics of nuclear monopole resonances \cite{chen}. 

\begin{figure}[tbp]
\center
 \includegraphics[width=0.745\textwidth]{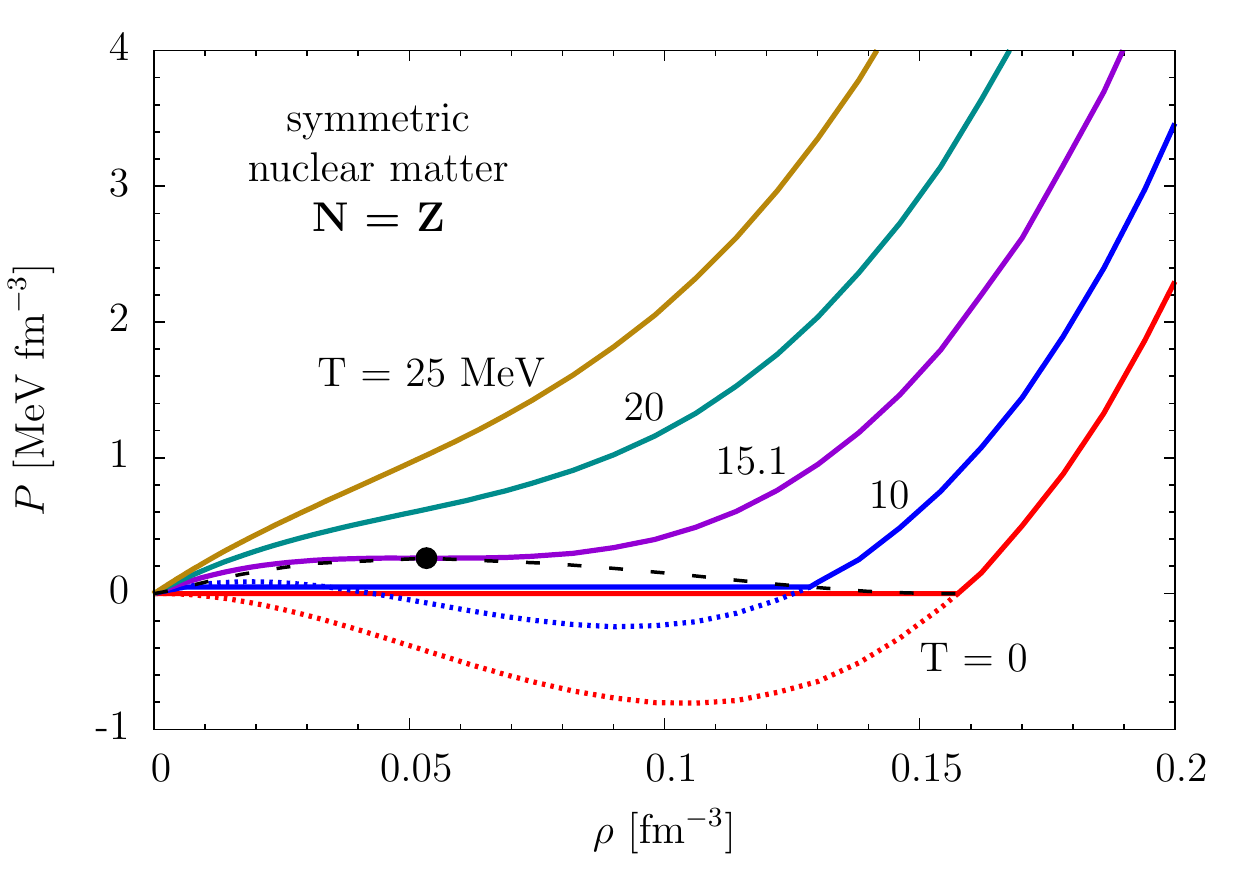}  
 \caption{Pressure isotherms as a function of density for symmetric nuclear matter displaying a first-order liquid-gas phase transition. The dotted lines at low temperature show the non-physical behaviour of the isotherms in the transition region. The physical pressure is calculated using the Maxwell construction. The dashed line delimits the boundary of the coexistence region. The dot indicates the critical pint ($ T_c \simeq 15.1 $ MeV, $ \rho_c \simeq \rho_0/3 $).} 
 \label{eos_50}
 \end{figure}

Fig.~\ref{eos_50} shows isotherms of the pressure $ P (\rho, T) $ (see Eq.~\eqref{pressure}) as a function of density $ \rho $. The emerging picture is qualitatively reminiscent of a van der Waals gas with its generic liquid-gas first-order phase transition. Indeed, chiral $ NN $ dynamics generates intermediate range attractive interactions, such as $ 2\pi$-exchange with intermediate $ \Delta$-excitation, that display a characteristic $ e^{-2\,m_\pi r} / r^6 $ behaviour at intermediate (1-2 fm) distances (see Ref.~\cite{kaiser5}). Such mechanisms account already for roughly half of the attraction required to bind nuclear matter at $ T = 0 $. The liquid-gas phase transition is then the result of a sensitive balance between intermediate range attraction and short-range repulsion, the latter represented by contact terms encoding (non-perturbative) physics not resolved in detail at the range of Fermi momenta $ k_F $ relevant to the present study.

The critical temperature for the liquid-gas phase transition is found at $ T_c \simeq 15.1 $ MeV in this calculation. For $ T < T_c $ the usual Maxwell construction \footnote{We briefly remind the main steps of the Maxwell construction. We plot first the free energy per particle $ \bar{F}(\rho) $ as a function of the volume per particle $ v = 1/\rho $. The region with negative curvature is unphysical and is eliminated through the double tangent construction. We search for the two points of the curve with the same tangent. This tangent line is the physical free energy in the phase coexistence region and its negative slope gives the physical pressure.} is applied, keeping the pressure constant in the liquid-gas coexistence region. Empirical values of this transition temperature deduced from multifragmentation and fission measurements locate $ T_c $ between 15 and 20 MeV \cite{karnaukhov}. Phenomenological Skyrme model studies \cite{sauer} gave a similar range for $ T_c $.

In order to assess possible uncertainties of the thermodynamic properties in our calculation we have considered variations of the parameters $ B_5 $ and $ \zeta $. Within the ranges $ -0.5\leq B_5\leq 0.5 $ and $ -0.8 \leq \zeta \leq -0.7 $ the critical temperature varies from $ T_c = 14.6 $ MeV to $ T_c = 15.8 $ MeV, i.e. this variation implies at most a change of $ \pm 5\% $. 

\begin{figure}[htbp]
\center
\includegraphics[width=0.6\textwidth]{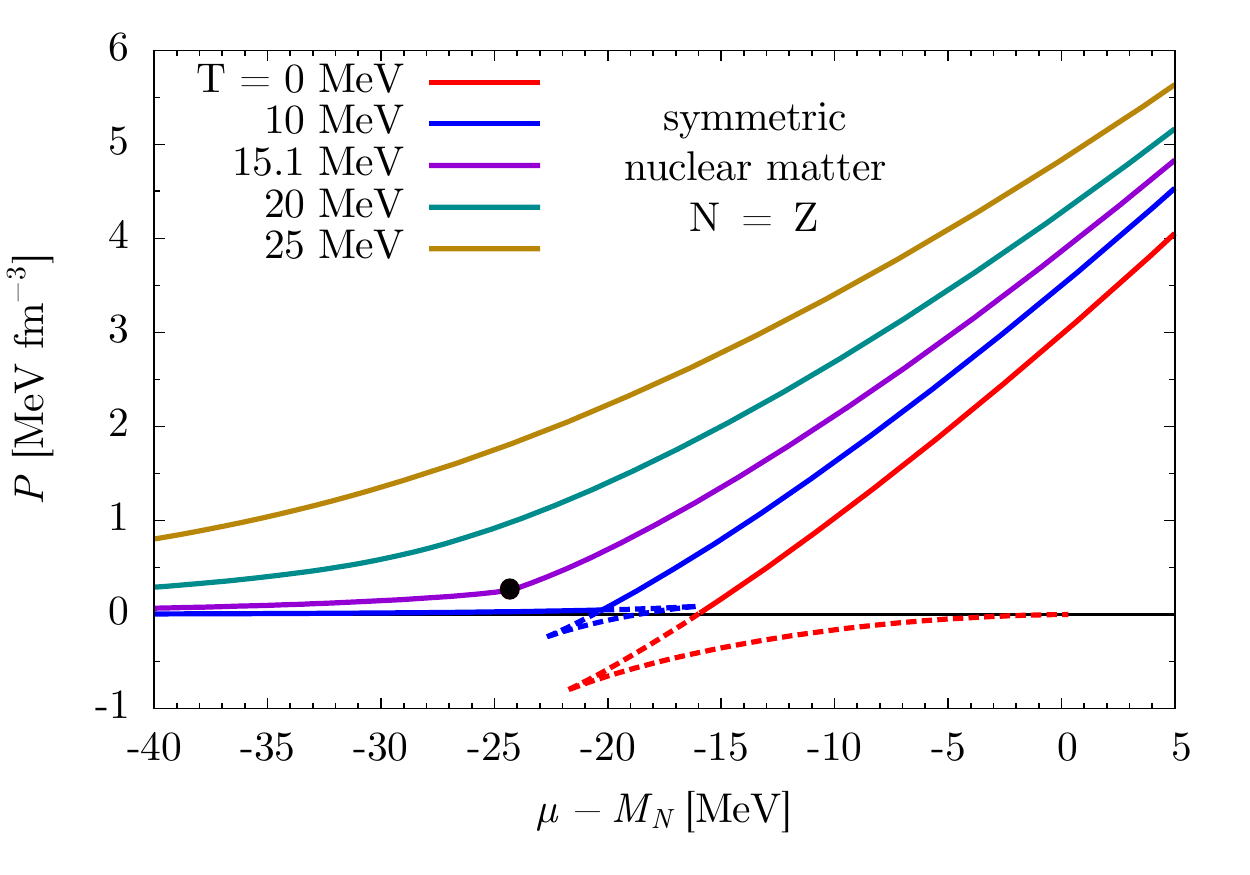}  
 \caption{Pressure isotherms as functions of the nucleon chemical potential $ \mu $ for isospin-symmetric nuclear matter. The dotted double-valued region of the curves at temperatures below $ T_c \simeq 15.1 $ MeV corresponds to the non-physical behaviour of the equation of state in the liquid-gas coexistence region. In this region the actual pressure and chemical potential are constant and determined by the Maxwell construction. The dot indicates the critical point.} 
 \label{pchem_50}
 \end{figure}

The discussion of the equation of state is completed by displaying the pressure as a function of the nucleon chemical potential (including the free nucleon mass),
\begin{equation}
\mu = M_N + \left( 1+\rho\, \frac{\partial}{\partial \rho} \right) \bar{F}(\rho, T)  \ ,
\end{equation} 
in Fig.~\ref{pchem_50}. The non-physical (dotted) curves in $ P(\rho, T) $ of Fig.~\ref{eos_50} find their correspondence in the double-valued behaviour of $ P $ as function of $ \mu $ at temperatures below $ T_c $. In this coexistence region the actual pressure and chemical potential are constant and given by the intersection point of the double-valued loop according to the Maxwell construction.  The temperature at which this loop reduces to a point is the critical temperature $ T_c $.

 \begin{figure}[htbp]
\centering
 \subfloat[][Temperature vs chemical potential.]{\label{t_mu_sym}
\includegraphics[width=.485\columnwidth]{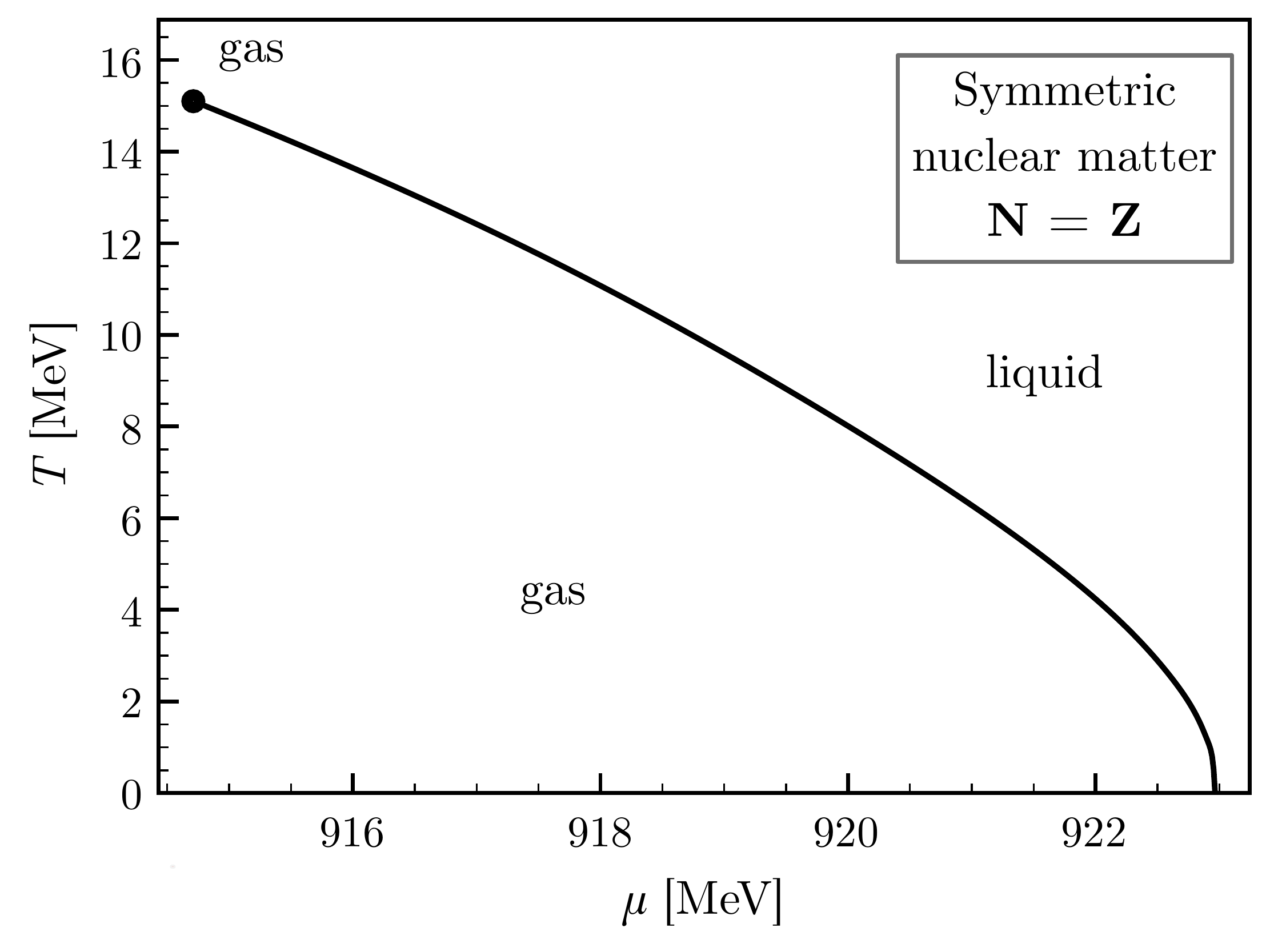}} \ \ 
\subfloat[][Pressure vs temperature.]{\label{p_t_sym}
\includegraphics[width=.485\columnwidth]{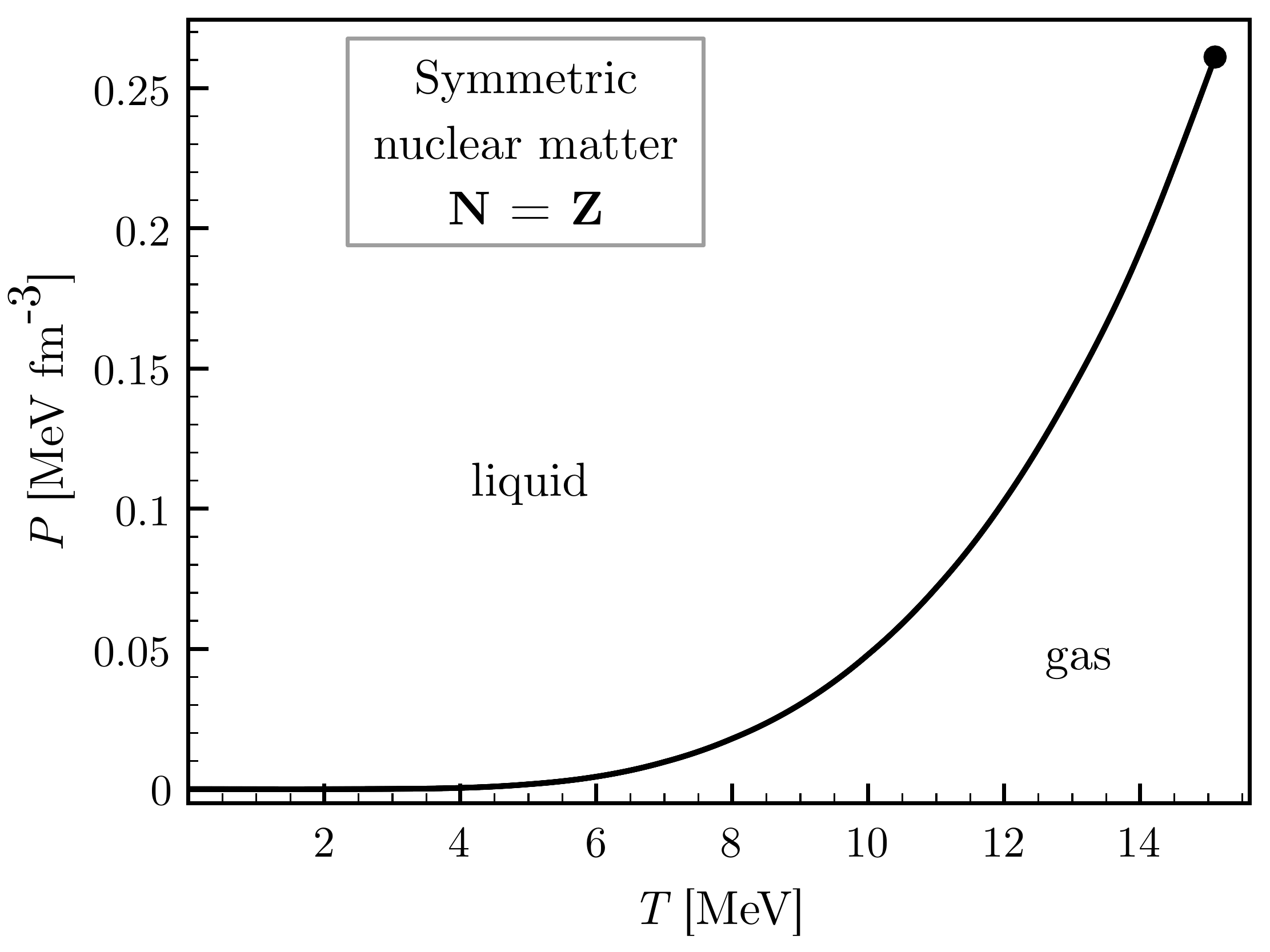}}   \\ 
 \subfloat[][Temperature vs density.]{\label{t_r_sym}
 \includegraphics[width=.58\columnwidth]{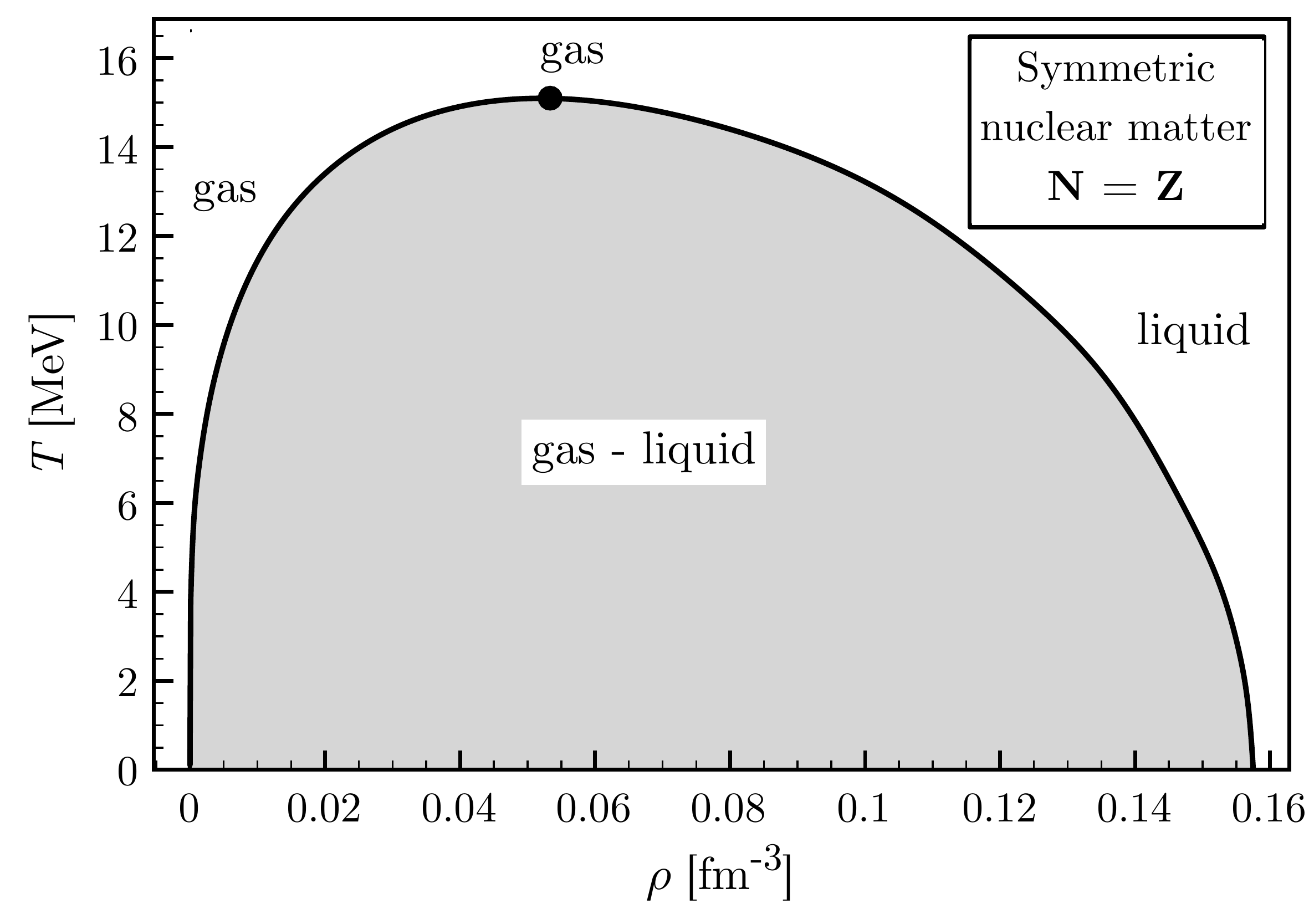}}
\caption{Phase diagrams of symmetric nuclear matter. The dot indicates the critical point.}
\label{phase_50}
\end{figure}

All relevant informations about the phase transition are  collected in the $ T - \mu $, $ P - T $ and $ T - \rho $ phase diagrams (Fig.~\ref{phase_50}). In Fig.~\ref{t_mu_sym} we display temperature versus chemical potential ($ T - \mu $) and in Fig.~\ref{p_t_sym}  pressure versus temperature  ($ P - T $). In these diagrams the coexistence region gets projected onto a first-order phase transition line. The first-order transition region terminates at the critical point indicated by the dot. The following critical values of thermodynamic quantities (pressure, baryon chemical potential and density) are found: $ P_c \simeq 0.261\, \text{MeV fm}^{-3} $, $ \mu_c \simeq 914.7 $ MeV and $ \rho_c \simeq 0.053\, \text{fm}^{-3} $ at $ T_c = 15.1 $ MeV. The $ T-\rho $ phase diagram, Fig.~\ref{t_r_sym}, clearly illustrates the extension of the gas-liquid coexistence region.

At $ T = 0 $ the third law of thermodynamics gives important constraints about the slope of the boundaries of the transition region. It implies that the boundaries of the transition region at $ T = 0 $ have an infinite slope in the $ T - \rho $ and $ T - \mu $ diagrams and a zero slope in the $ P - T $ diagram. Moreover, the chemical potential at $ T = 0 $ is known and given by the total energy per particle at the saturation point, i.e. $ \mu = M_N + \bar{E}_0 \simeq 923 $ MeV. 

\section{Equation of state of isospin-asymmetric nuclear matter} \label{asymmetric}
 
 \begin{table}[tbp]
\centering
\begin{tabular}{ c*{4}{c}}
\toprule
\textbf{Proton fraction} &  \multicolumn{4}{c}{\textit{Critical point}}  \\
 $ \mathbf{x_p} $ & $ T_c $  & $ P_c $ &  $ \mu_c $ & $ \rho_c $  \\
   &  [MeV]  & [MeV fm$^{-3}$] & [MeV]  & [fm$^{-3}$]   \\
   \midrule
  \bfseries 0.5       & 15.1 & 0.26 &  914.7  & 0.053	\\ 
  \bfseries 0.4     &	14.7 & 0.25 &   916.1   & 0.053	\\ 
  \bfseries 0.3   &   13.4   & 0.22 &   920.1  & 0.049	\\
 \bfseries 0.2   &  10.9 & 0.16 & 927.2	& 0.044	\\
 \bfseries 0.1   &  6.4 &  0.08 &  937.6 & 0.037	\\
 \bfseries 0.053   &	0 & 0.02 &943.6 & 0.032\\
\bottomrule
\end{tabular}
\caption{Critical point as a function of the proton fraction $ x_p = Z/A $:  listed are the value of temperature, pressure, chemical potential and density. }
 \label{critical}
 \end{table}
 
 \begin{figure}[htbp]
\centering
\subfloat[][]
{\includegraphics[width=.49\columnwidth]{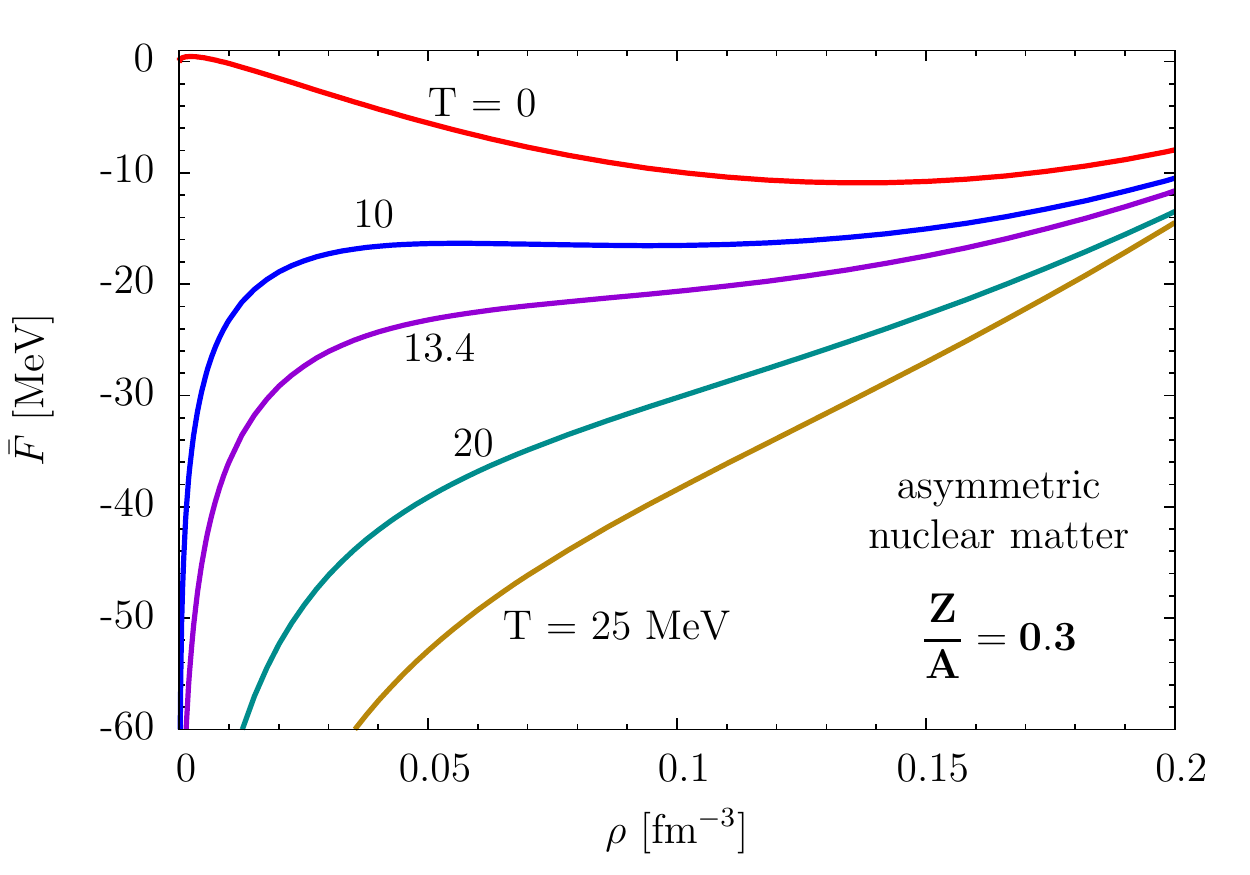}} \ 
\subfloat[][]
{\includegraphics[width=.49\columnwidth]{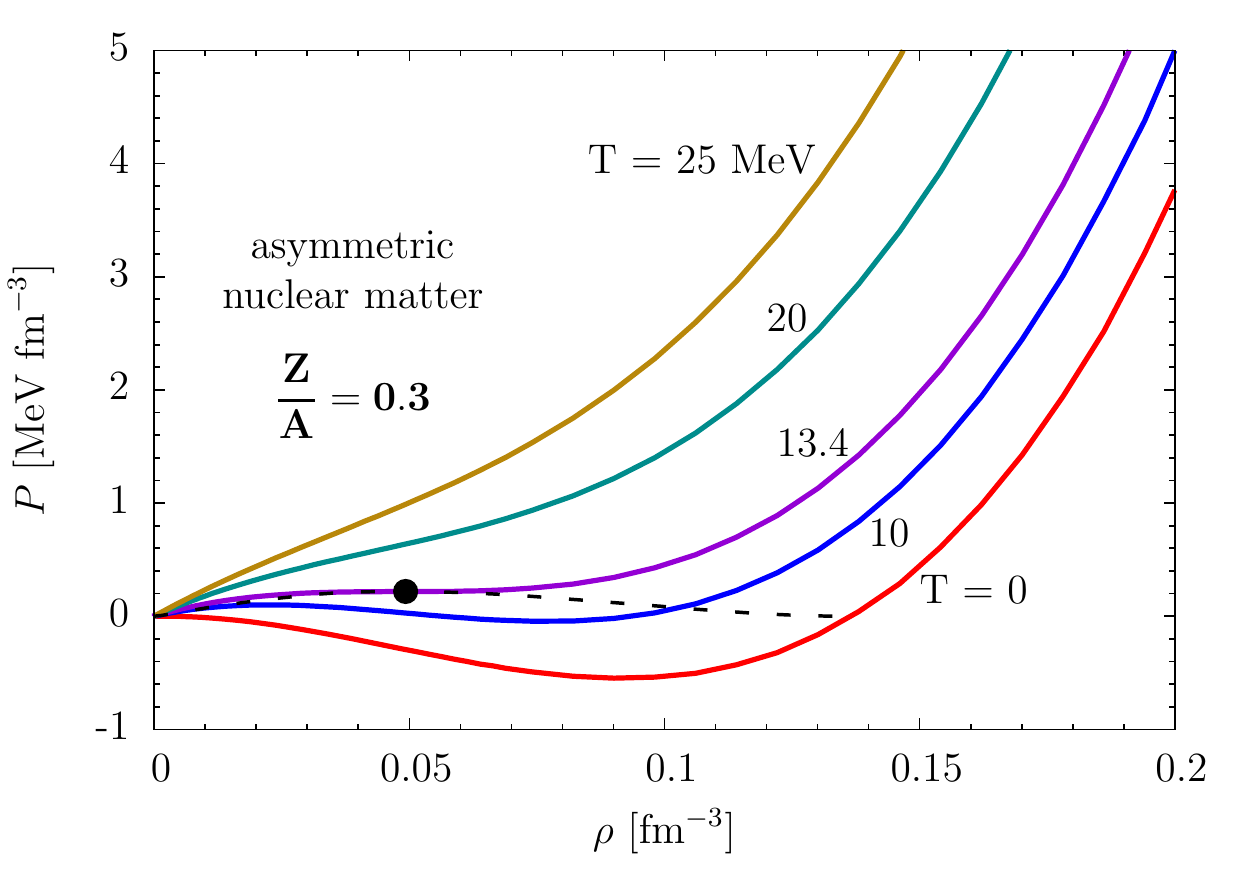}}   \\
 \subfloat[][]
{\includegraphics[width=.49\columnwidth]{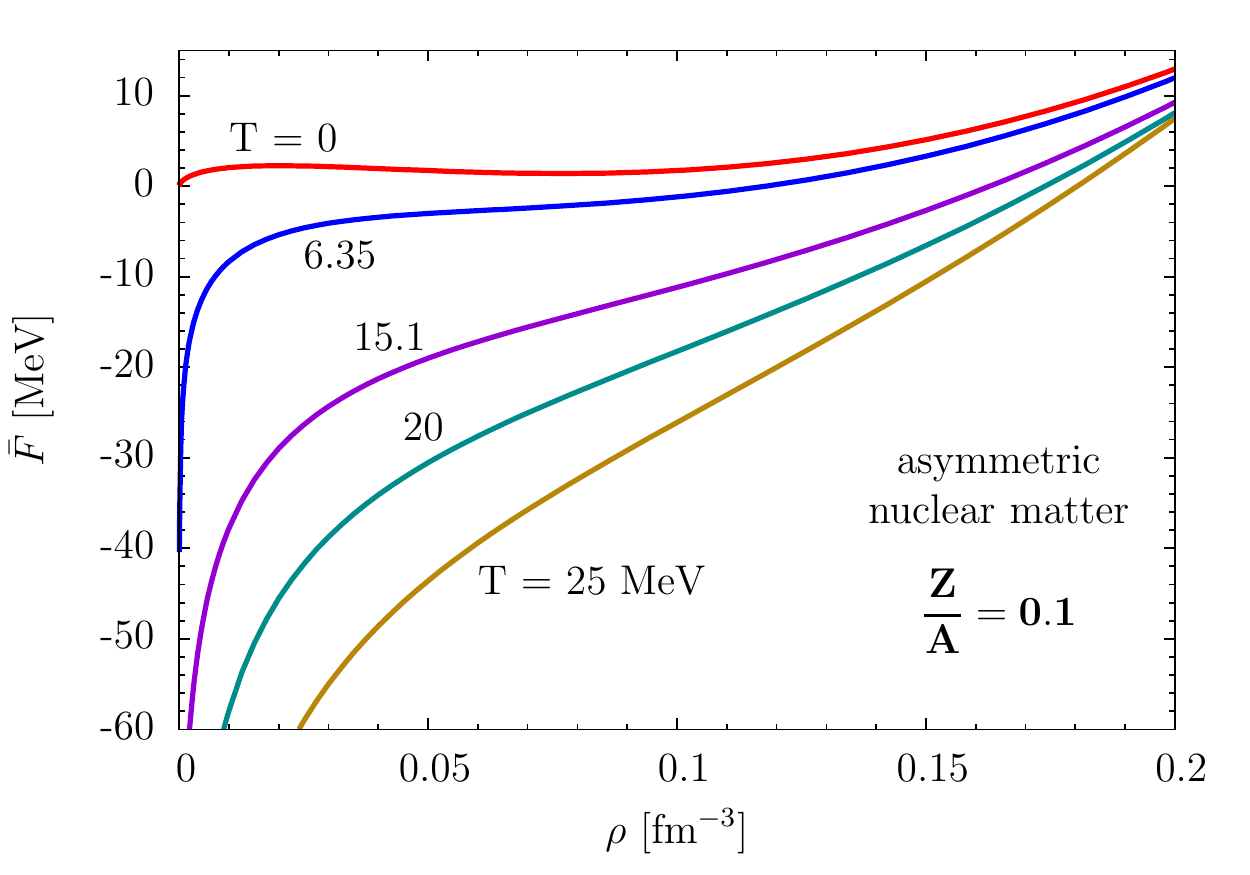}} \ 
\subfloat[][]
{\includegraphics[width=.49\columnwidth]{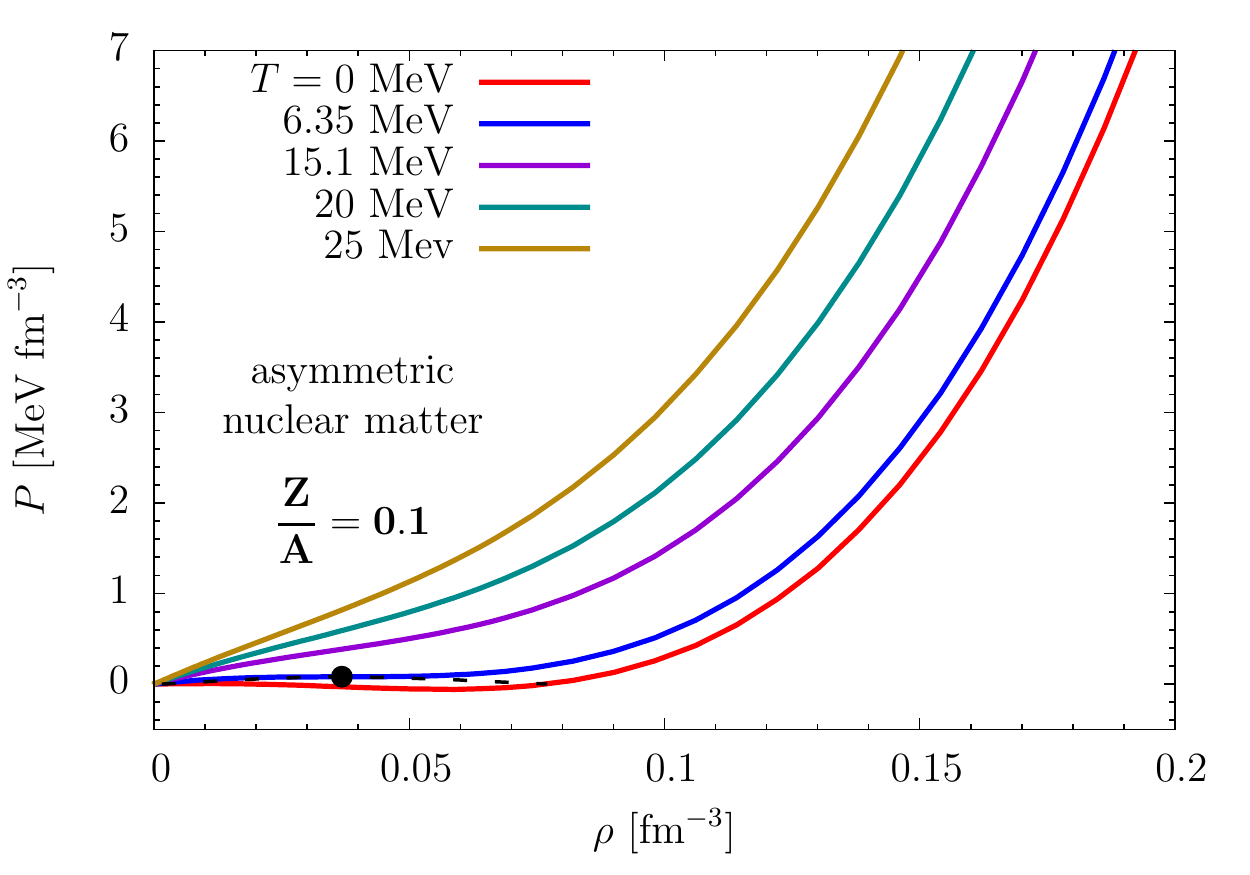}}   \\
 \subfloat[][]
{\includegraphics[width=.49\columnwidth]{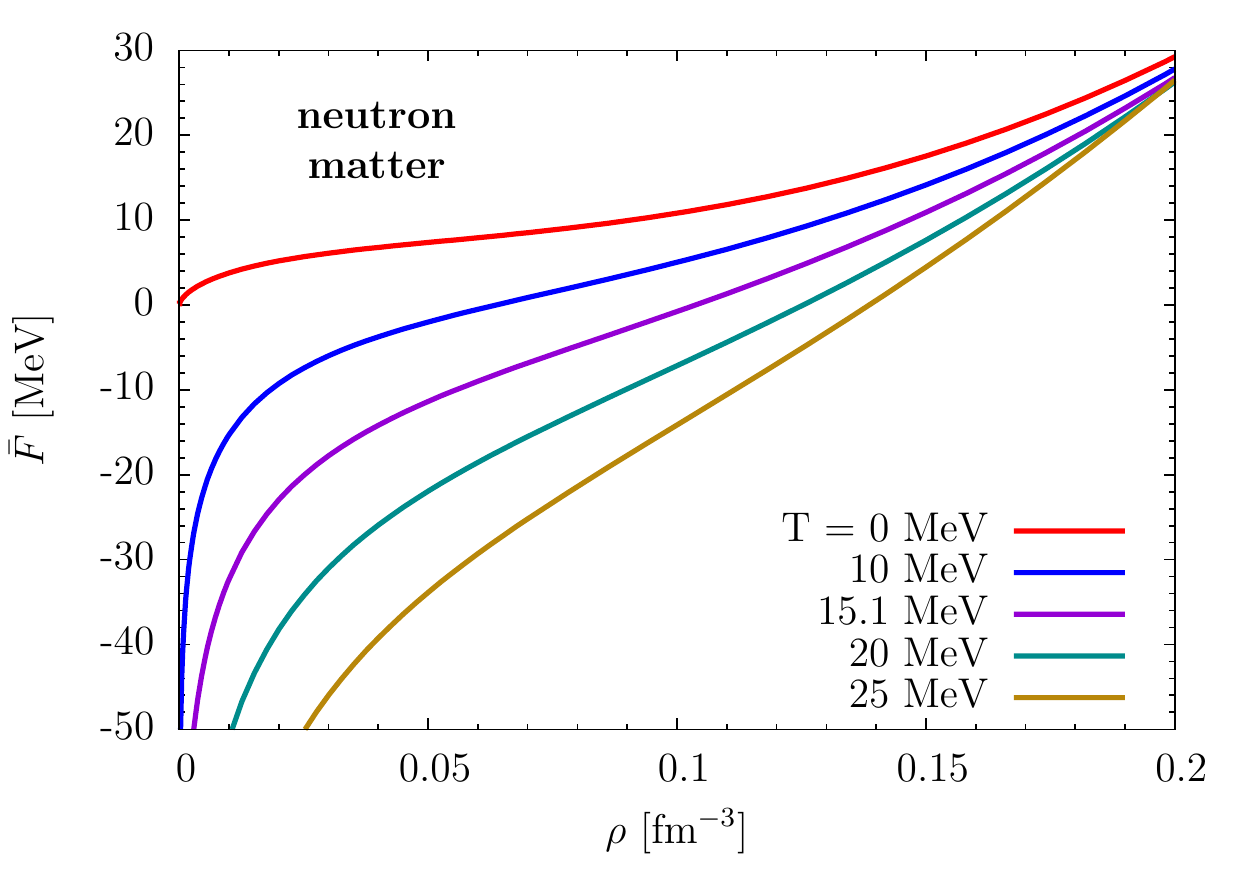}} \ 
\subfloat[][]
{\includegraphics[width=.49\columnwidth]{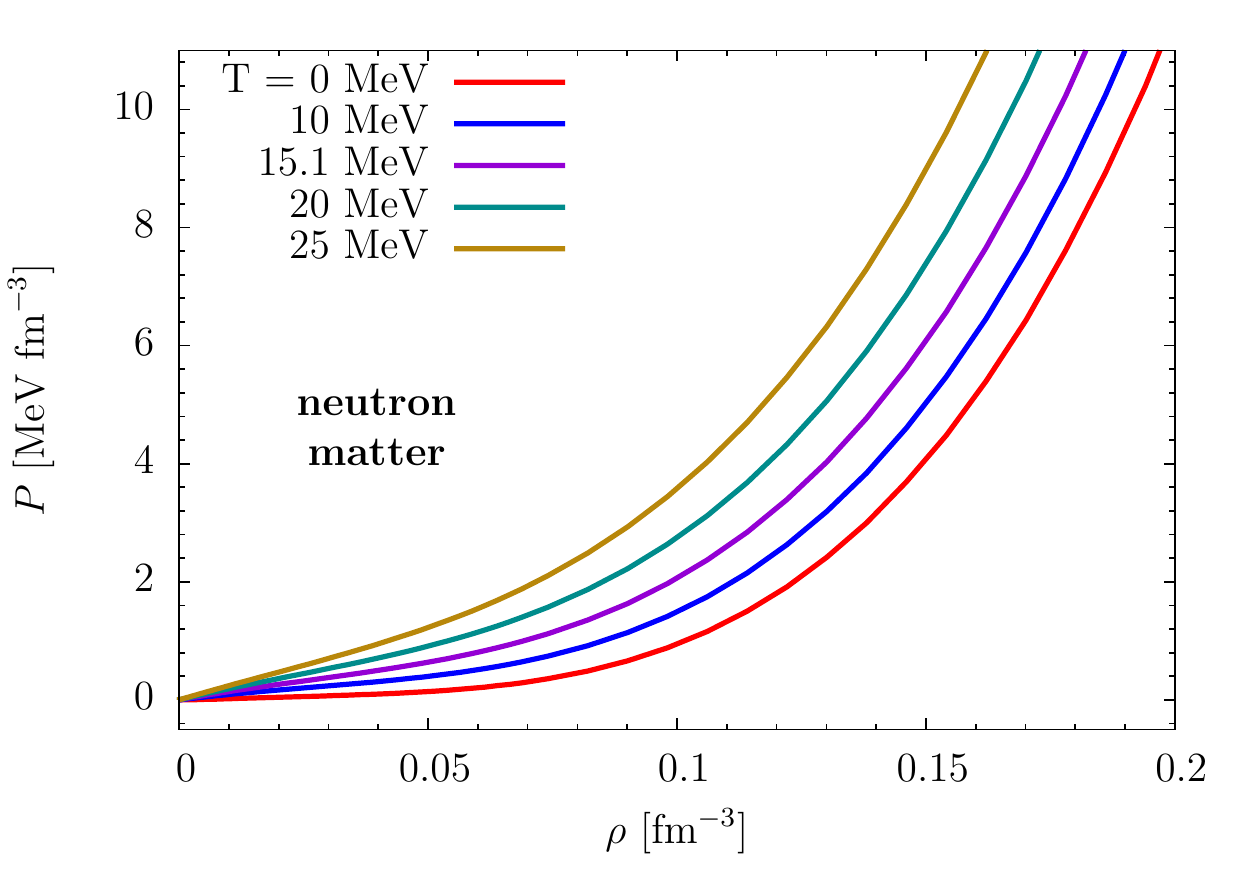}}   
 \caption{Isospin-asymmetric nuclear matter: free energy per particle (left column) and pressure isotherms (right column) as a function of nucleon density for proton fractions $ x_p $ = 0.3, 0.1 and 0. The coexistence region, with boundary delimited by the dashed line, gradually diminishes with increasing isospin-asymmetry until it disappears at $ x_p = 0.053 $. The dot indicates the critical point.}
\label{eos_asy}
\end{figure}
 
In this section we investigate the dependence of nuclear matter properties on the isospin-asymmetry. The amount of asymmetry is given in terms of the proton fraction, $ x_p = Z/A = \rho_p/(\rho_p+\rho_n) $. The calculations are again based on the expansion of the free energy density, Eq.~\eqref{convolution}, with input specified in \ref{appB}.

The resulting equations of state for different values of the proton fraction, $ x_p $ = 0.3, 0.1 and 0, are shown in Fig.~\ref{eos_asy}. The limiting case $ x_p = 0 $ corresponds to pure neutron matter. The left column of Fig.~\ref{eos_asy} displays the free energy per particle, the right column shows the pressure, both as a function of the baryon density $ \rho = \rho_n + \rho_p $. 

As the neutron-proton asymmetry increases, the free energy and the pressure increase at given density, indicating the reduced binding in neutron-rich matter. Pure neutron matter is unbound. The systematics displayed in Fig.~\ref{eos_asy} is almost entirely controlled by the isospin dependence of the one- and two-pion exchange forces  once the relevant contact terms are fixed to yield the empirical asymmetry energy at $ T=0 $, about 34 MeV. The coexistence region of the liquid-gas phase transition shrinks with decreasing proton fraction until it disappears and only  the (interacting) Fermi gas phase remains.

\begin{figure}[tbp]
\centering
{\includegraphics[width=.65\columnwidth]{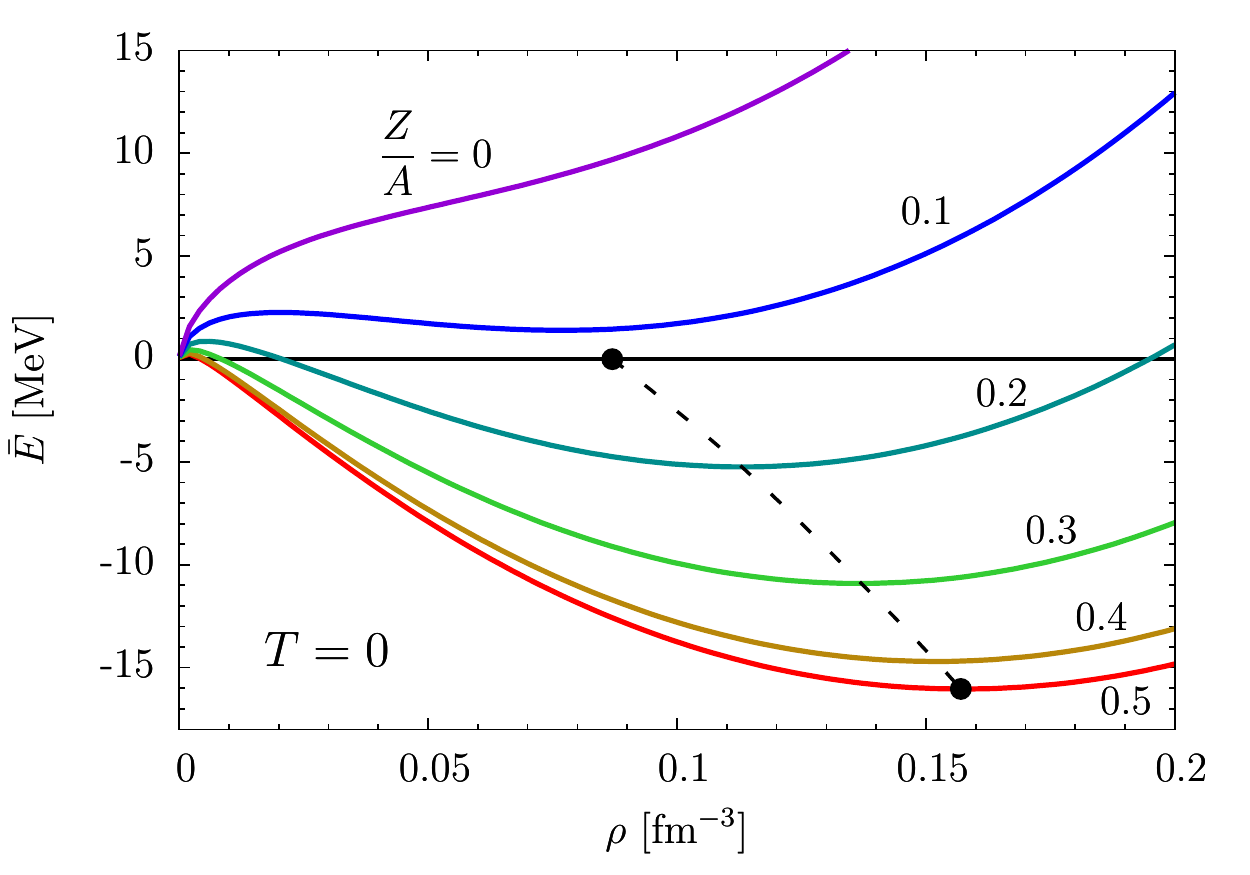}} 
\caption{Dependence of energy per particle and saturation point of nuclear matter at $ T = 0 $ on the asymmetry. The solid lines represent the energy per particle  as a function of nucleon density $ \rho = \rho_n+ \rho_p $ for different proton fractions $ x_p = Z/A $. The dashed line shows the trajectory of the saturation point as $ x_p $ varies. For $ x_p \lesssim 0.12 $ the energy is always positive.}
\label{sat_point}
\end{figure}

To complete the picture,  consider the evolution of the saturation point, defined as the minimum of the energy curve at $ T = 0 $, in Fig.~\ref{sat_point}. Starting from its minimum for symmetric nuclear matter, $ \bar{E}_0 \simeq -16 $ MeV at $ \rho_0 \simeq 0.157\,\text{fm}^{-3} $, the binding energy per nucleon is reduced continuously with decreasing $ x_p $ until it vanishes at a proton fraction $ x_p \simeq 0.12 $. Beyond this point neutron-rich matter at $ T = 0 $ is unbound.    
 
 \begin{figure}[tbp]
\centering
 \subfloat[][Temperature vs chemical potential.]
{\includegraphics[width=.49\columnwidth]{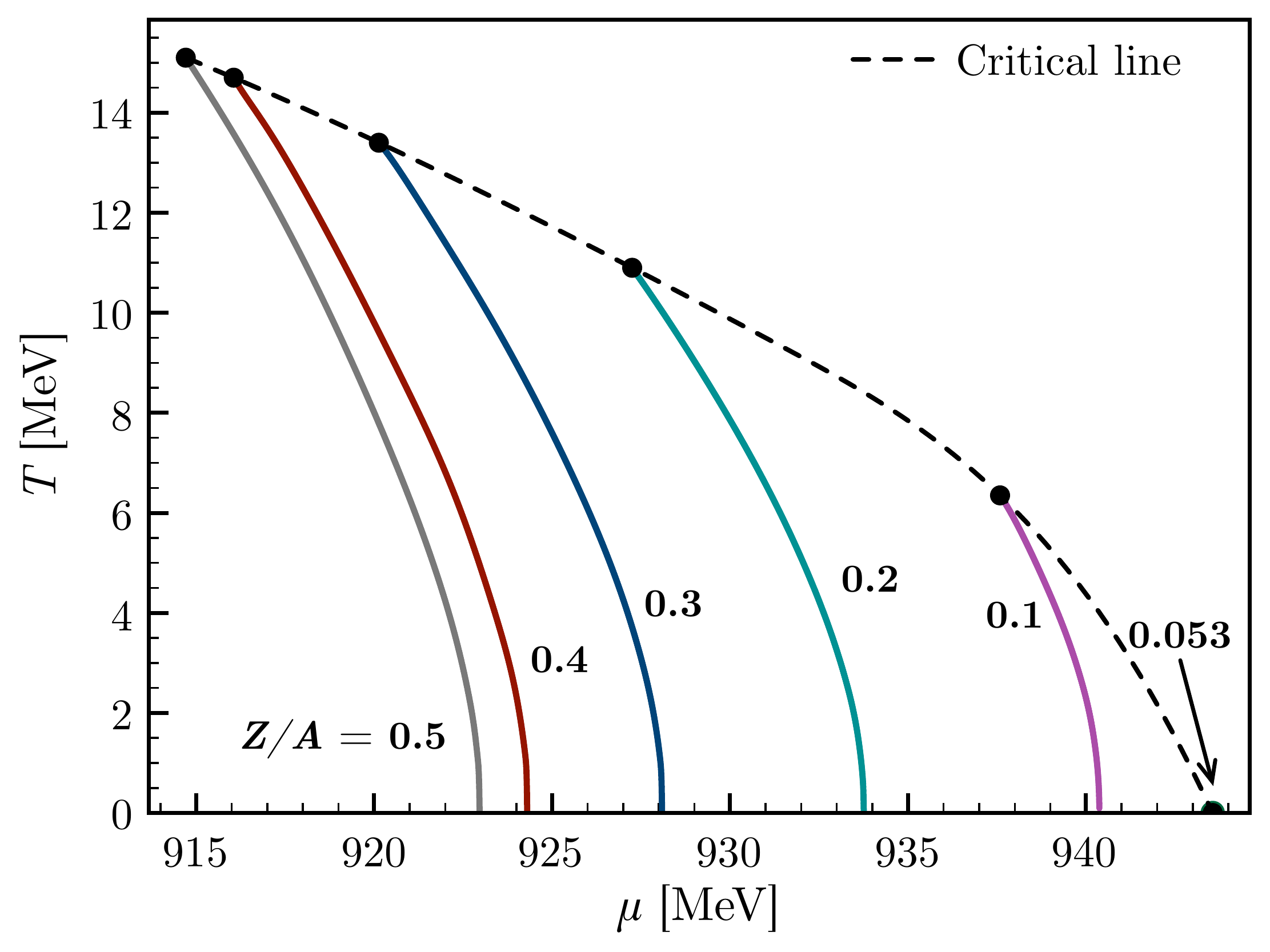}} \ \ 
\subfloat[][Pressure vs temperature.]
{\includegraphics[width=.49\columnwidth]{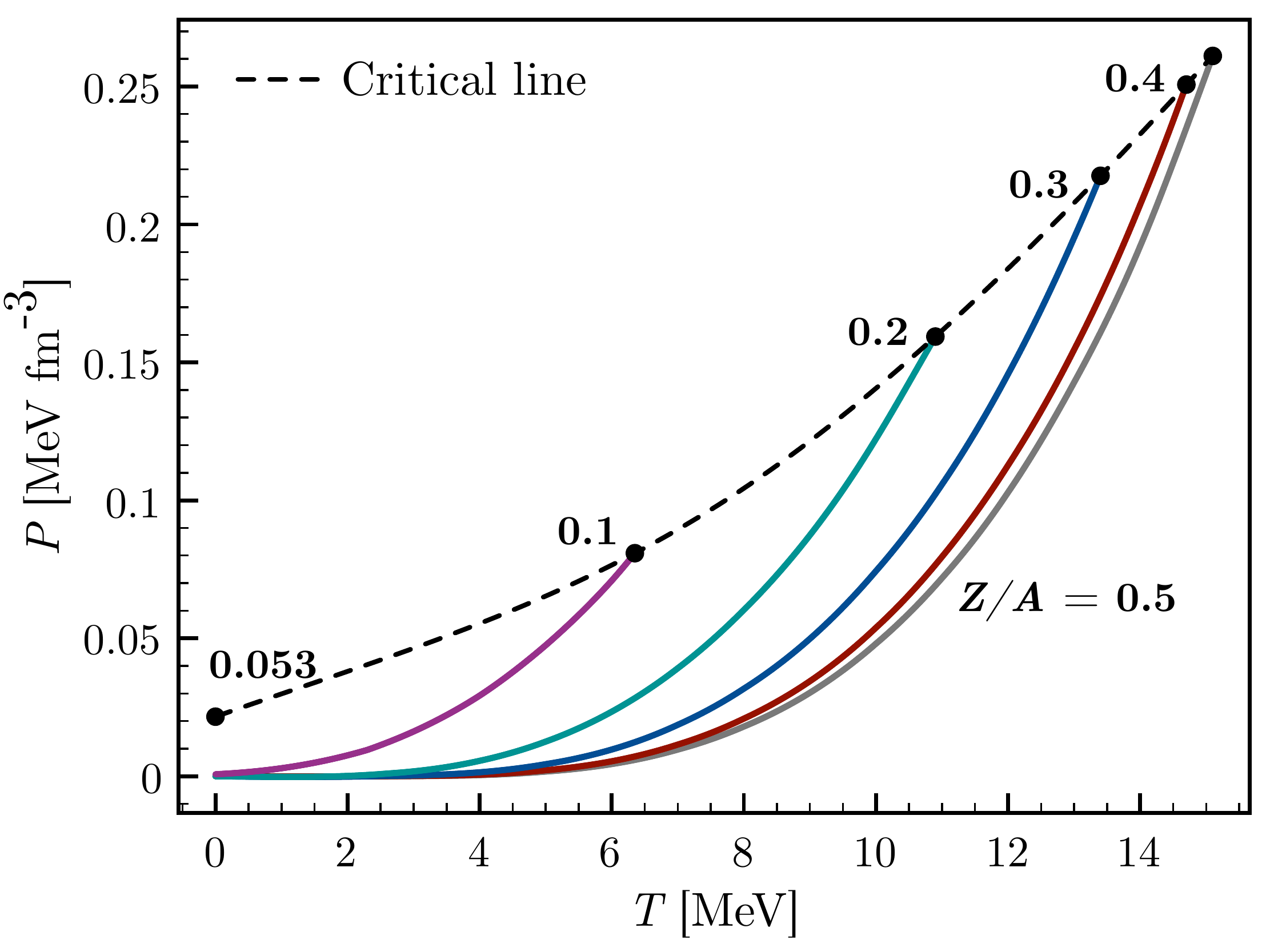}}   \\ 
 \subfloat[][Temperature vs density.]
 {\includegraphics[width=.8\columnwidth]{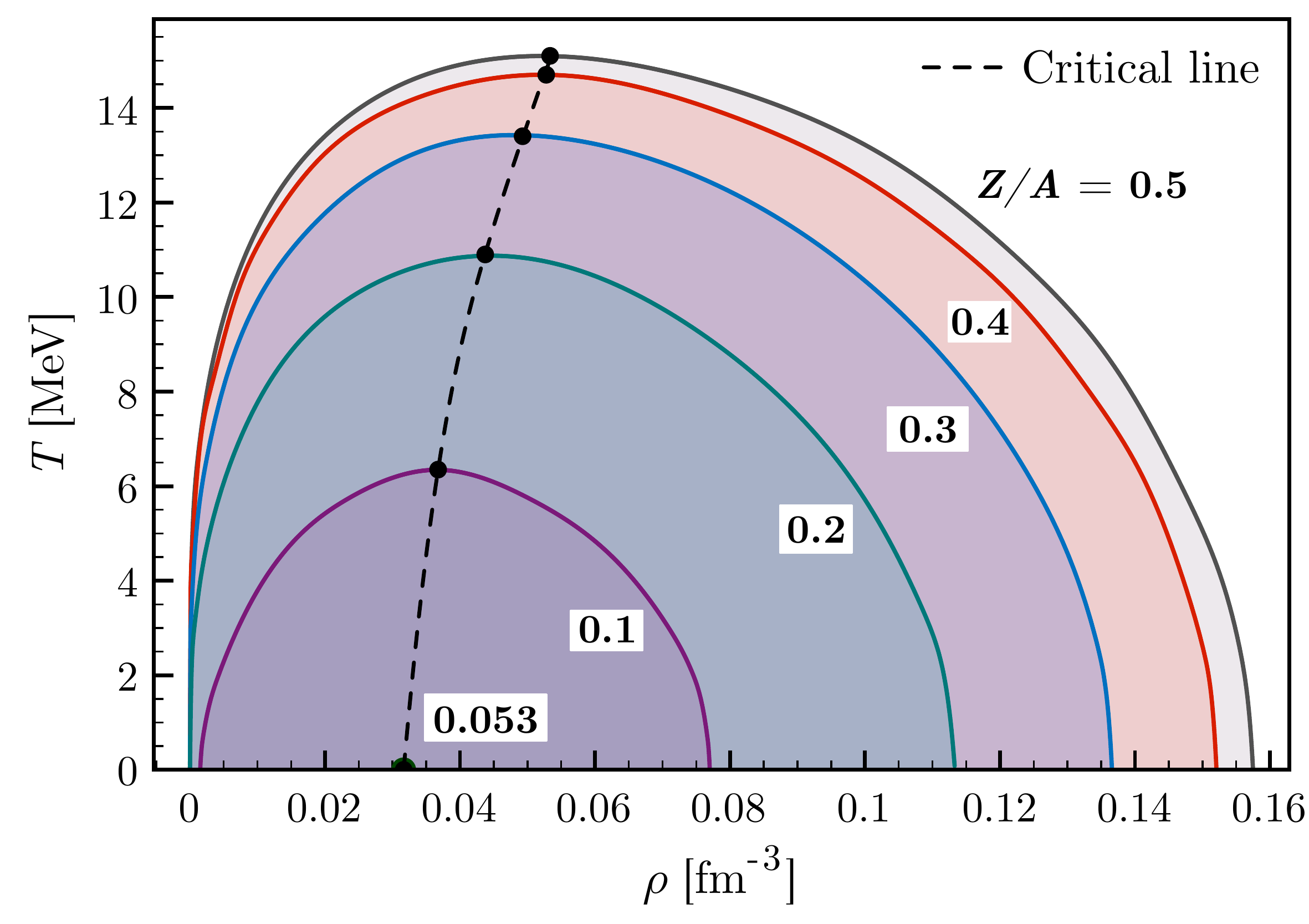}}
\caption{Phase diagrams of isospin-asymmetric nuclear matter for different proton fractions. The dashed line shows  the evolution of the critical point.}
\label{phase_asy}
\end{figure}

Phase diagrams of isospin-asymmetric nuclear matter are shown in Fig.~\ref{phase_asy} for different values of the proton fraction, demonstrating how the matter evolves with increasing asymmetry. The dashed line shows the evolution of the critical point and its disappearance at $ x_p \simeq 0.053 $. At this particular proton fraction the coexistence region reduces to a point, as can be seen in the $ T - \rho $ diagram, meaning that the liquid-gas phase transition does not take place any more. Neutron-rich matter with $ x_p \lesssim 0.053 $ is always in a gaseous phase. The values of the thermodynamic quantities at the different critical points are listed in Tab.~\ref{critical}. Note that the critical point disappears at a small but finite pressure.
  
The following features of the phase diagram in Fig.~\ref{phase_asy} are worth noting. Above a proton fraction close to $ x_p \simeq 0.1 $, the gas-liquid coexistence region starts at zero density. The $ T = 0 $ boundary at which this coexistence turns into a Fermi liquid covers a density range between about $ 0.5\, \rho_0 $ and normal nuclear matter density $ \rho_0 $, as one moves from $ x_p = 0.1$ to 0.5. A qualitative change in the behaviour of the coexistence region takes place at $ x_p = 0.12 $, the point at which the binding energy at $ T=0 $ vanishes. Following the dashed line in Fig.~\ref{sat_point} that represents absolute minima of the energy per particle, one observes that for $ x_p\lesssim 0.12 $ there is still a local minimum in $ \bar{E}(\rho, x_p) $, but the absolute minimum is now located at $ \rho = 0 $. Consequently, neutron-rich nuclear matter in the vicinity of $ x_p \lesssim 0.12 $ is a gas at very low density and $ T=0 $, and then enters the coexistence region as the density increases. In the range $ 0.053 \lesssim x_p \lesssim 0.12 $ nuclear matter is not self-bound but it can still have a liquid-gas phase transition.

The present chiral thermodynamics framework for nuclear matter is, of course, oversimplified at low densities where nuclear clustering takes place. A detailed study combining the appearance of light (deuteron, triton and helium) clusters with relativistic mean field phenomenology \cite{typel} suggests, however, only modest changes of the $ T-\mu $ phase diagram, such as a shift in the position of the critical point with changes of less than $ 10 \% $ in $ T_c $ and less than $ 1 \% $ in $ \mu_c $, when cluster formation is incorporated.

\section{Asymmetry free energy} \label{asymmetry}
 
Introducing the asymmetry parameter  
\begin{equation}
 \delta = (\rho_n - \rho_p) / \rho = 1-2x_p \ ,
 \end{equation} 
 the free energy per particle can be expanded in powers of $ \delta $ around the free energy of isospin-symmetric nuclear matter:
 \begin{equation} \label{as_fit2}
 	\bar{F}(\rho_p,\rho_n, T) = \bar{F}(\rho, T) + A(\rho, T)\, \delta^2 + \mathcal{O} (\delta^4) \ .
 \end{equation}
 This defines the asymmetry free energy per particle, $ A(\rho, T) $. The expansion of $ \bar{F} $ involves only even powers of $ \delta $ as long as we ignore isospin-symmetry breaking effects. In this limit nuclear matter is invariant under the interchange of protons and neutrons.
 
 \begin{figure}[tbp]
\centering
{\includegraphics[width=.49\columnwidth]{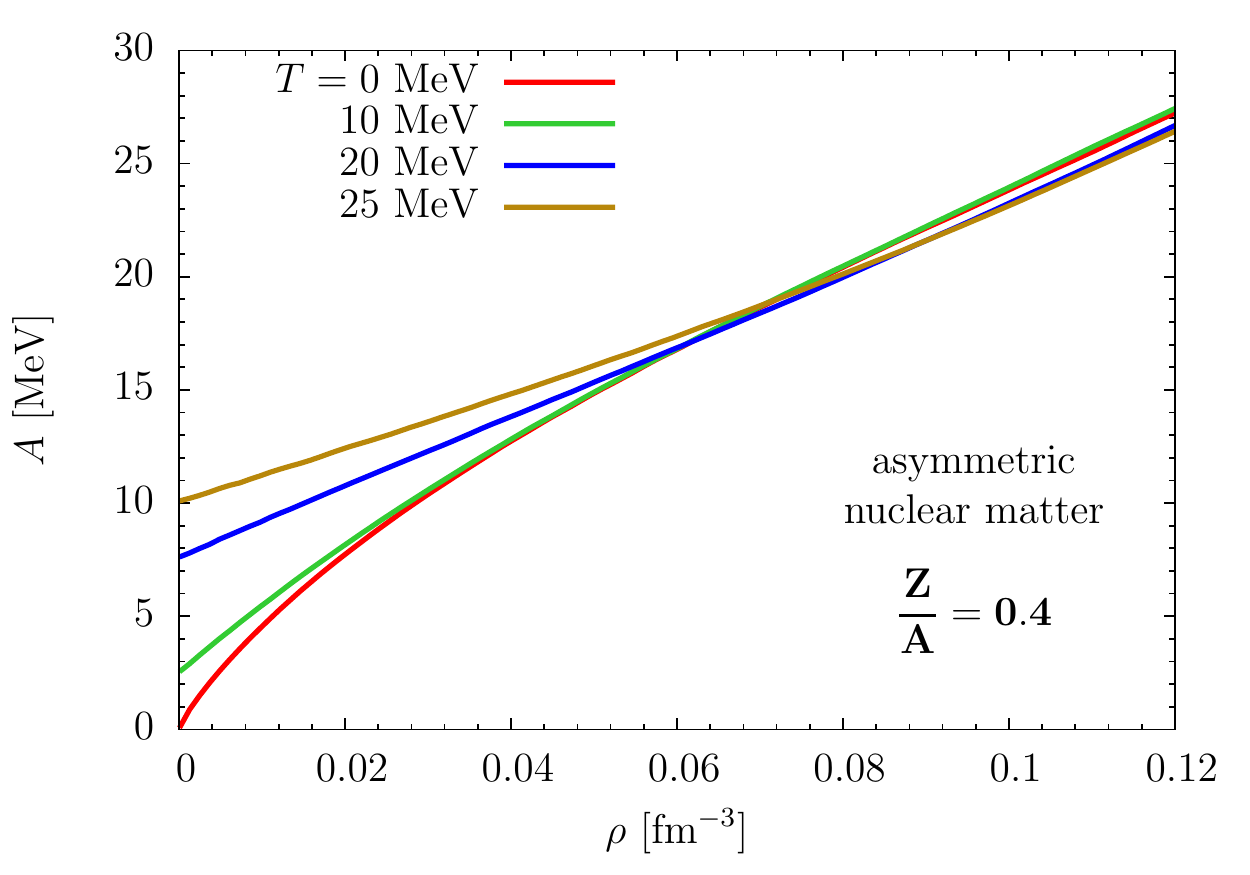}} 
{\includegraphics[width=.49\columnwidth]{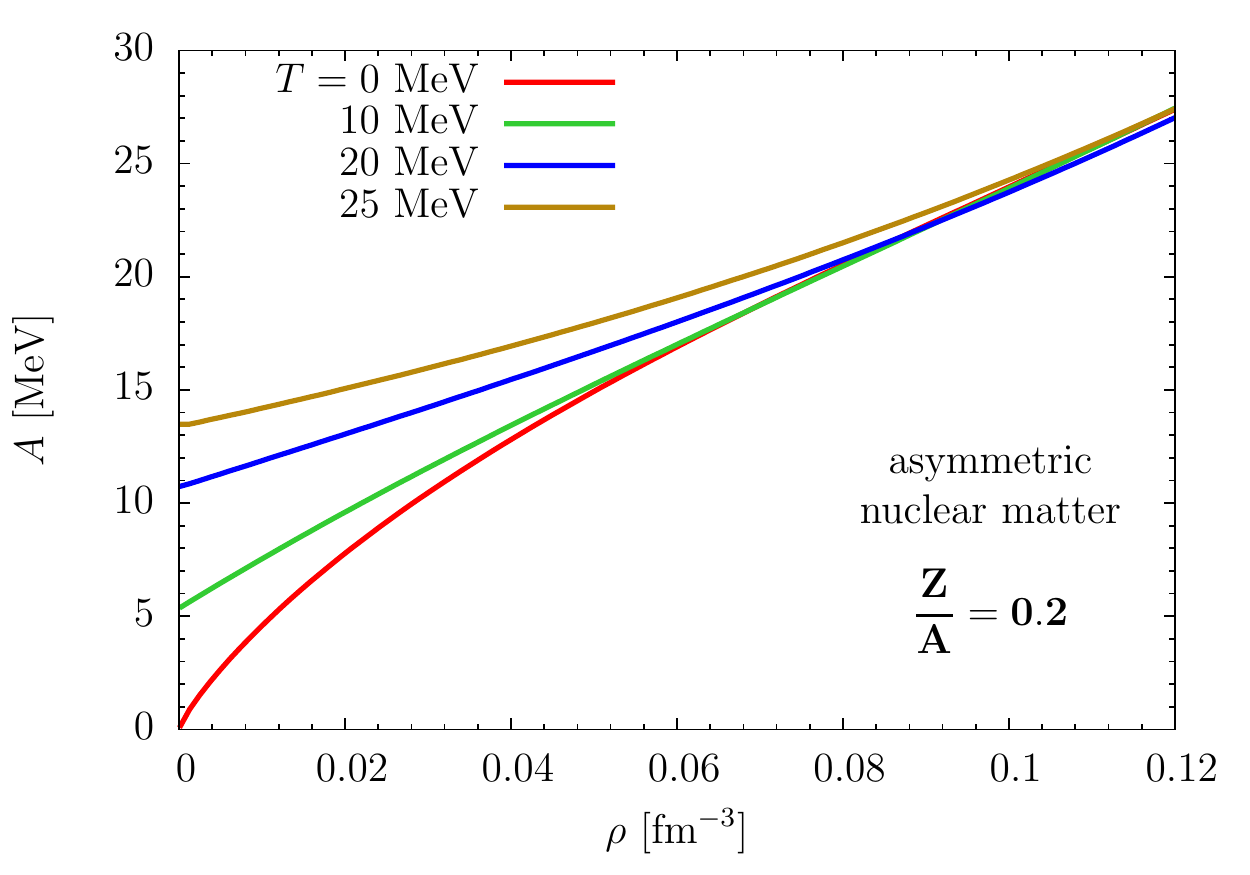}}   
\caption{Asymmetry free energy per particle as a function of the density for different temperatures. In the left  panel $ \delta = 0.2 $ ($ x_p = 0.4 $), in the right panel $ \delta = 0.6 $ ($ x_p = 0.2 $).}
\label{asyenergy}
\end{figure}
 
 In Fig.~\ref{asyenergy} we show the behaviour of the asymmetry free energy $ A(\rho, T) $ for $ \delta = 0.2 $ ($ x_p = 0.4 $) and $ \delta = 0.6 $ ($ x_p = 0.2 $) as a function of the nucleon density for different temperatures. We note that the asymmetry free energy is sensitive to the temperature only at low densities, $ \rho< 0.1\,\text{fm}^{-3} $.   
 
\begin{figure}[tbp]
\centering
{\includegraphics[width=.49\columnwidth]{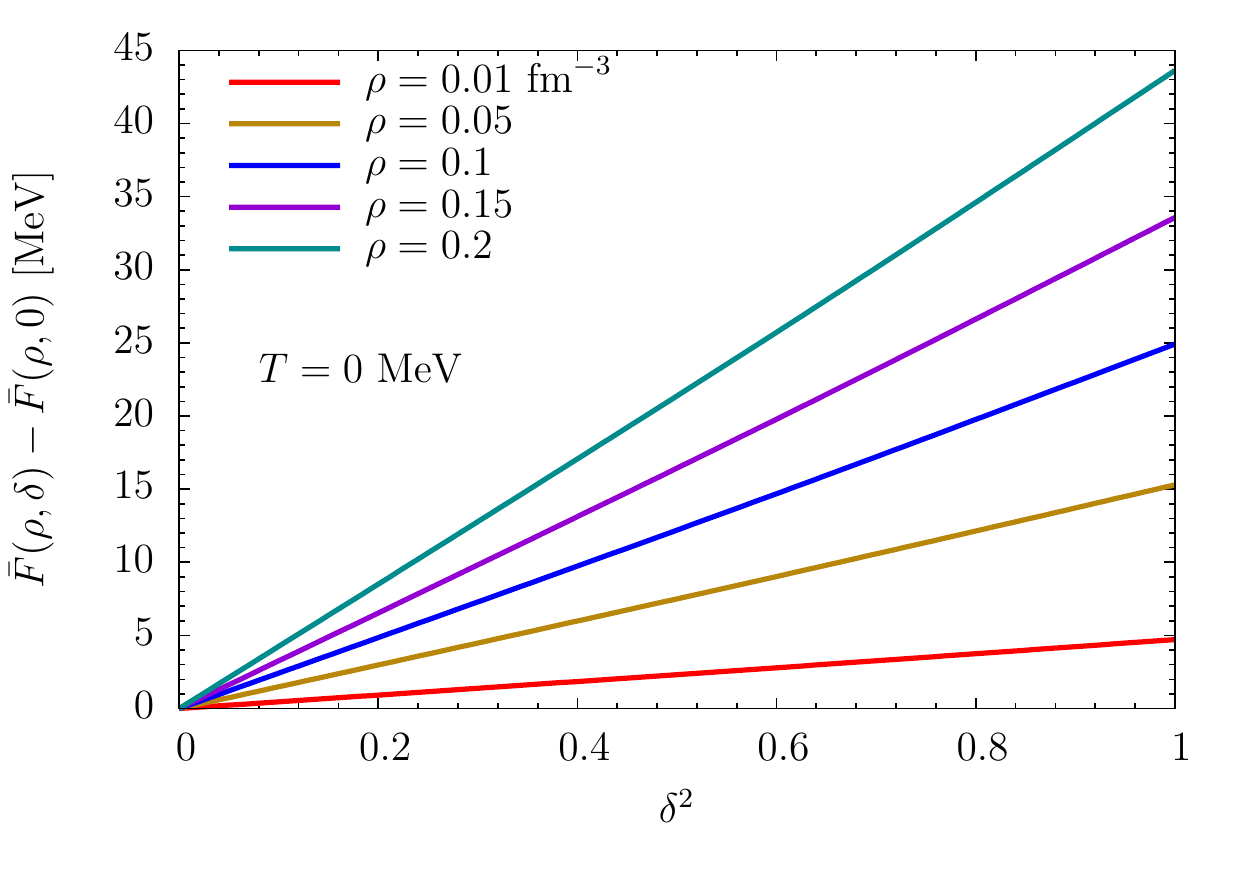}} 
{\includegraphics[width=.49\columnwidth]{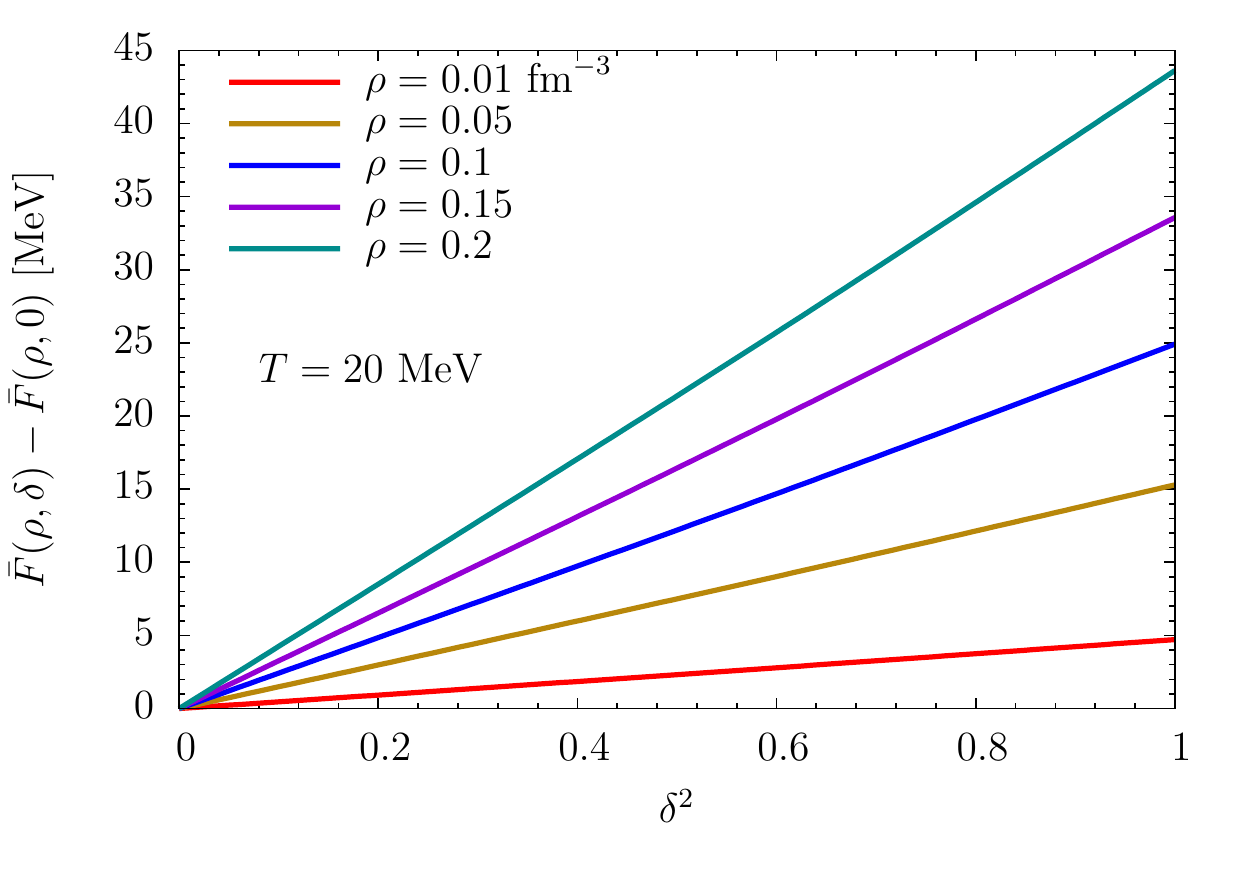}}   
\caption{Free energy per particle of nuclear matter as a function of the asymmetry parameter $ \delta^2 $ for different densities.}
\label{as_fit}
\end{figure}

 A test of the validity of the parabolic approximation \label{f_as} is  shown in Fig.~\ref{as_fit}, where we plot the free energy difference with respect to isospin-symmetric nuclear matter as a function of $ \delta^2 $ for different densities $ \rho = \rho_n + \rho_p $. At $ T = 0 $ (left plot of Fig.~\ref{as_fit}) the linear dependence on $ \delta^2 $ is seen to be realized very well even up to large $ \delta $. At higher temperature ($ T = 20 $ MeV, right plot), a slight bending is observed especially at low density. In summary, Eq.~\eqref{as_fit2} is confirmed to be a good approximation of the free energy; the term of order $ \delta^4 $ is generally negligible for most applications, even for large values of $ \delta $.  This feature has also been observed in other calculations \cite{baoanli,bombaci}. Estimates give a value of the quartic term smaller than 1 MeV at the saturation point \cite{baoanli}. 
   
  At the saturation density $ \rho_0 $  we have imposed $ A(\rho_0, T = 0) \simeq 34.0 $ MeV in our calculation in order to fix the contact terms associated with the isospin-dependent part of the interaction. For comparison, a relativistic mean-field model \cite{vretenar} constrained by the properties of selected nuclei gives $ A(\rho_0) = 34 \pm 2 $ MeV. A more recent estimate using the same relativistic mean-field phenomenology constrained by giant dipole resonances \cite{cao}  suggests $ A(\rho_0) $ in the range (33 - 37) MeV. A further estimate in the same paper gives the asymmetry energy at lower density, $ \rho = 0.1 $ fm$^{-3} $, in a window between 21 and 23 MeV,  slightly lower than our calculated value $ A(0.1\, \mbox{fm}^{-3}, T = 0) \simeq 23.9 $ MeV. Previous determinations from extensive fits of nuclide masses \cite{blaizot,seeger} gave $ A(\rho_0) $ between 33  and 37 MeV. 

Around the saturation point $ \rho_0 $, the asymmetry energy at $ T = 0 $ can be expanded in powers of $ \rho-\rho_0 $ as follows:
\begin{equation}
	A(\rho) = A(\rho_0) + L\,\frac{\rho - \rho_0}{3\,\rho_0} + \frac{K_{as}}{2} \left(  \frac{\rho - \rho_0}{3\,\rho_0} \right)^2 + \dots  
\end{equation}
We extract the coefficients $ L \simeq 90.1 $ MeV and $ K_{as} \simeq 153 $ MeV. The value of $ L $, in particular, is compatible with empirical constraints from isospin diffusion which give $ L = 88 \pm 25 $ MeV \cite{baoanli}. 

For small asymmetries $ \delta $, the saturation density gets lowered to $ \rho_0 (1 - 3L\,\delta^2/K ) $ and the corresponding  compression modulus $ K(\delta) $ is often expressed as an expansion in powers of $ \delta $:
\begin{equation}
	K(\delta) = K + K_\tau \delta^2 + \mathcal{O}(\delta^4) \ , \qquad K_\tau= K_{as} - 6\,L \ ,
\end{equation}
where $ K $ is the compressibility of symmetric nuclear matter and $ K_\tau $ is called isobaric compressibility. 
Our calculated value is $ K_\tau = -388 $ MeV. The empirical determination suffers from large uncertainties. A recent result from measurements of the isotopic dependence of giant  monopole resonances in even-A isotopes  gives $ K_\tau = -550 \pm 100 $ MeV \cite{li}.

\section{Chiral four-body contributions}\label{sec5}

In this section we study four-body contributions to the nuclear matter equation of state. We restrict ourselves to the evaluation of a particularly simple class of four-loop in-medium diagrams (at $ T=0 $). The leading long-range four-nucleon interaction is constructed by connecting the four nucleon lines through exchanged pions which couple together according to the chiral pion-pion interaction \cite{4npot}. However, since the off-shell chiral $ 4\pi$-vertex is involved in the process this contribution alone is representation dependent and thus not unique. It has to be supplemented by the four-nucleon interaction generated additionally by the chiral  $ NN 3\pi$-vertex, where the three pions emitted from one nucleon are absorbed on each of the other three nucleons. The combination of contributions from the chiral $ \pi\pi$-interaction and the chiral $ NN 3\pi$-vertex is representation independent and thus gives rise to unique and physically meaningful results. The (isovector) Weinberg-Tomozawa  $ NN\pi\pi $ involves the energies of the exchanged pions which in the present application are differences of nucleon kinetic energies. The corresponding four-nucleon interaction is then a relativistic $1/M_N^2$-correction. Note that we do not follow the method of unitary transformations of Ref.~\cite{4npot}. In this scheme three-body and two-body forces and even disconnected diagrams give rise to induced four-body forces.  We consider here only diagrams related to ``genuine'' four-nucleon forces.

\begin{figure}[tbp] 
 \center
 \includegraphics[height=2.4cm]{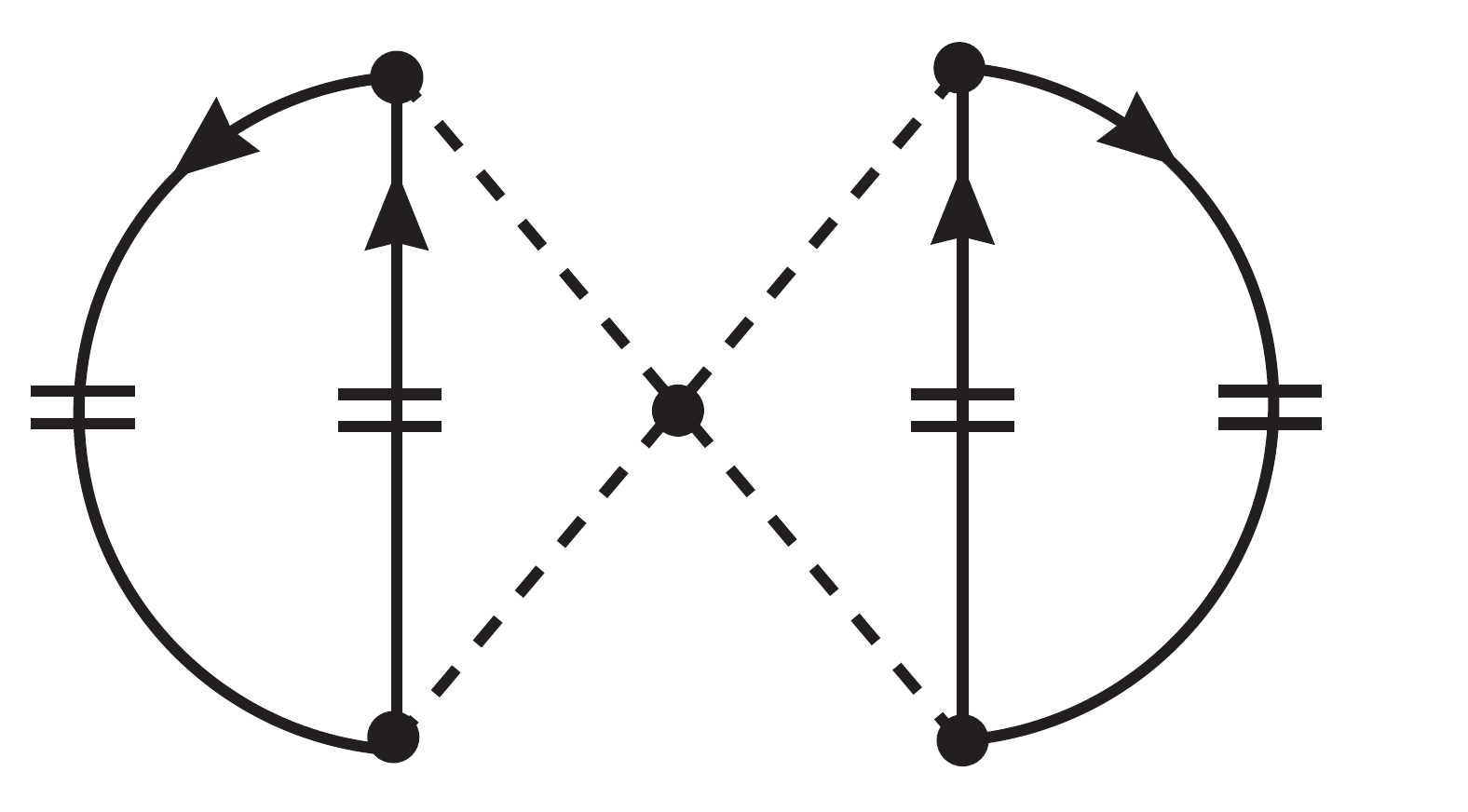}     \quad
  \includegraphics[height=2.4cm]{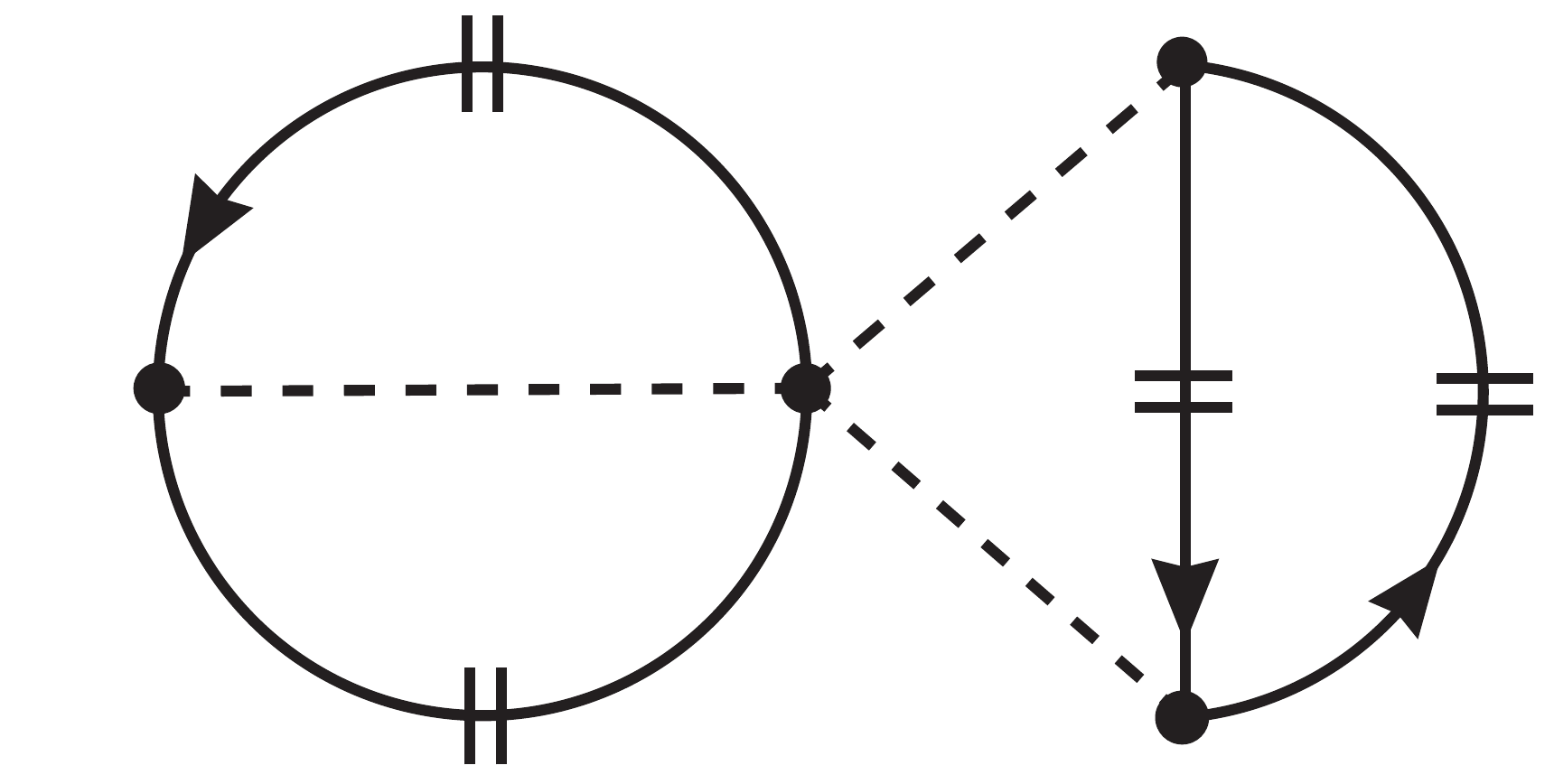}     
\caption{In-medium Hartree diagrams related to the leading-order chiral four-nucleon interaction. Their combinatoric factors are $ 1/8 $ and $ 1/2 $. The four-medium insertions are indicated by the double-line breaks on the nucleon propagators.}
\label{4nucleon}
\end{figure}

\begin{figure}[htbp]
\begin{center}
\includegraphics[scale=0.4,clip]{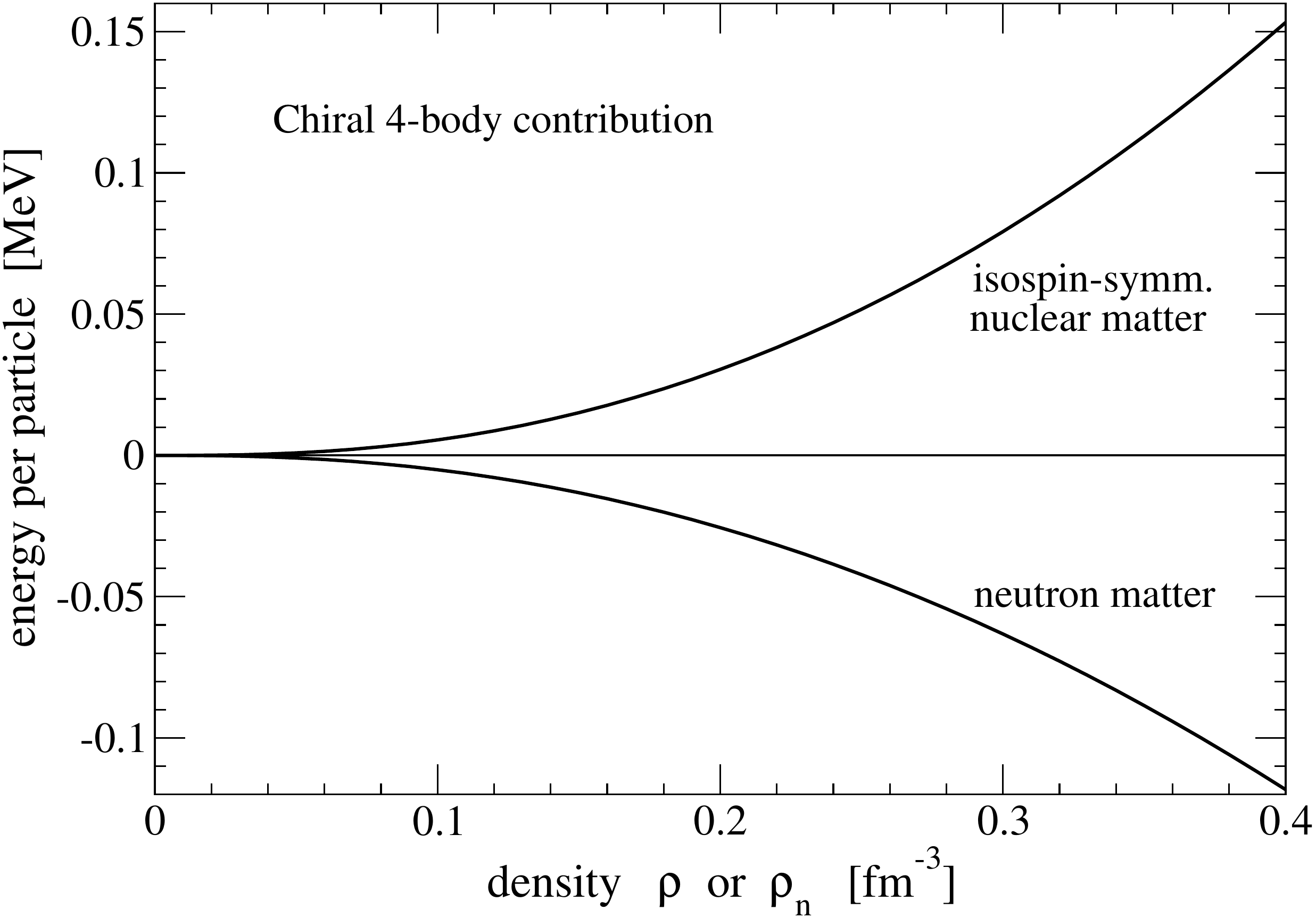}
\end{center}
\caption{Contribution to the energy per particle arising from the long-range
chiral four-nucleon interaction. Upper curve: isospin-symmetric nuclear matter,
lower curve: pure neutron matter.}
\label{4energy}
\end{figure}

Fig.~\ref{4nucleon} shows the in-medium diagrams one obtains by closing the four nucleon lines to two rings. The combinatoric factors of these diagrams are $1/8$ and $1/2$, respectively. Diagrams with more rings are trivially zero, due to a vanishing spin-trace. The integrals over the Fermi spheres factorize and thus can be readily solved. After adding both diagrams one gets the following Hartree contribution from the chiral four-nucleon interaction to the energy per particle of isospin-symmetric nuclear matter:  
   \begin{equation} \label{4body}
\bar E(\rho)^{(4N)}= {9g_A^4m_\pi ^7 u\over (4\pi f_\pi)^6} \bigg[u^2
-{1\over 2}-2u \arctan 2u+\bigg(1+{1\over 8u^2}\bigg) \ln(1+4u^2)\bigg]^2 \,,
\end{equation}
with $u = k_F/m_\pi$. Note that after adding both diagrams in Fig.~\ref{4nucleon} the only remainder of the chiral $\pi\pi$-interaction is a constant factor $-3m_\pi^2/ f_\pi^2$. This feature has ultimately lead to the expression with a complete square in Eq.~\eqref{4body}. The upper curve in Fig.~\ref{4energy} shows the repulsive four-body
contribution $\bar E(\rho)^{(4N)}$ as a function of the nucleon density $\rho = 2k_F^3/3\pi^2$. One observes that up to twice normal nuclear matter density, $2\rho_0 = 0.32\,$fm$^{-3}$, the effects from the genuine long-range (pion-induced) four-nucleon correlations stay below $0.1\,$MeV, and  thus are negligibly small. To be specific, at saturation density one has $\bar E(\rho_0)^{(4N)}\simeq 18\,$keV.

The chiral four-body contribution in pure neutron matter is also of interest. In this case only neutral pions are present which leads to a reduced isospin weight factor. Doing the calculation, one finds for the energy per particle
$ \bar E_n(\rho_n)^{(4N)} $ of pure neutron matter the same expression as in Eq.\eqref{4body} with an additional prefactor $-1/6$ and the substitution $ u = k_n/m_\pi$, with $k_n$ the neutron Fermi momentum.  The lower curve in Fig.~\ref{4energy} shows the attractive four-body contribution $\bar E_n(\rho_n)^{(4N)}$ as a function of the neutron density $\rho_n = k_n^3/3\pi^2$. Note that the increase of the Fermi momentum (by a factor $2^{1/3}$) has effectively compensated the (isospin) reduction factor $1/6$. Again, with less than $0.1\,$MeV in magnitude this
long-range four-neutron correlations are negligibly small. In addition to the two Hartree diagrams shown in Fig.~\ref{4nucleon} there are (exchange) Fock diagrams with one single closed nucleon ring. We do not enter here into the
intricate numerical evaluation of these (non-factorizable) four-loop diagrams. But from general considerations one knows that Fock terms come with opposite sign and reduced spin- and isospin weight factors, and therefore one
expects some even smaller and partly compensating effects.


A more comprehensive study of pion-induced four-nucleon forces including also virtual $\Delta(1232)$-isobar excitations in the context of nuclear matter calculations is an ongoing project. In comparison to eq,~\eqref{4body} one expects more significant effects from these mechanisms. First, there exist three-ring diagrams with higher spin-isospin weight factors and, secondly, the corresponding contributions to the energy per particle do not vanish in the chiral limit $ m_{\pi} \to 0 $. Note that Ref.~\cite{4body_delta} has already studied such pion-induced $\Delta(1232)$-excitation mechanisms for the four-nucleon system.

\section{Comparison to other approaches}

In the recent literature several calculations of the thermodynamic properties of nuclear matter using different approaches can be found \cite{rios,vidana,soma,rios2,wu}. In the following we comment on some of these, based either on the self-consistent Green's function (SCGF) method \cite{soma} or on mean field theory \cite{wu}.

As pointed out in Ref.~\cite{rios}, many traditional studies of nuclear matter at finite temperature in mean field approximation treat  the temperature dependence in a naive way. The temperature dependence arises mainly from the replacement of  the step function momentum distribution at zero temperature by the corresponding Fermi-Dirac  distribution. In this way the temperature dependence of the phenomenological interactions accounting for the correlations between nucleons is ignored. On the other hand, in many-body calculations based on microscopic approaches, such as ours, the nuclear medium generates Pauli blocking effects that are weakened with increasing  temperature. Consequently nuclear matter properties and correlations in the medium are temperature dependent in a non-trivial way. 

In the SCGF approach the ladder approximation provides the minimal scheme for the description of the effective $ NN $ interaction in the medium. The three-body Green's function is decomposed into a one-body and a two-body propagator and arranged in an integral equation for the in-medium $ T$-matrix that sums the ladder diagrams to all orders. The set of equations including the $ T$-matrix, the nucleon self-energy and the single-particle Green's  function can then be solved self-consistently. In Ref.~\cite{soma} the equation of state of isospin-symmetric nuclear matter has been calculated using such an approach in combination with the CD-Bonn and Nijmegen $ NN $ potentials. Contributions of three body forces are included via an effective (density-dependent) two-body interaction. The pressure $ P(\rho, T, \delta) $ is found to be strongly dependent on three body correlations and the liquid-gas coexistence region gets reduced in size when they are included. With the CD-Bonn $ NN $ potential the critical temperature results in $ T_c = 12.5 $ MeV. When using the Nijmegen potential $ T_c $ reduces to 11.5 MeV. The critical density lies in the range $ \rho_c\simeq 0.09 - 0.11\,\text{fm}^{-3} $ and the critical pressure is about $ 0.15 \, \text{MeV fm}^{-3} $. The results of these calculations, including the very low value of $ T_c $,  differs significantly from ours (compare with Tab.~\ref{critical}).

In Ref.~\cite{wu} the properties of nuclear matter have been calculated in the framework of relativistic mean field theory with density-dependent meson-nucleon couplings accounting for medium modifications. This method has been successful in describing many properties of both nuclear matter and finite nuclei. For symmetric nuclear matter a critical temperature of $ T_c = 13.2 $ MeV is found, while the liquid-gas coexistence region vanishes at $ x_p \simeq 0.07 $. The emerging picture is closer to the one we have obtained in the present paper. The main difference is a somewhat smaller phase transition region. We recall that the empirical range for the critical temperature is $ T_c \simeq (15 - 20) $ MeV \cite{karnaukhov}.

\section{Summary} \label{summary}

The present work extends previous calculations of nuclear chiral thermodynamics from symmetric nuclear matter to the isospin-asymmetric case. A virial expansion of the free-energy density is treated to three-loop order using in-medium chiral perturbation theory. One- and two-pion exchange processes in the nuclear medium are computed explicitly with inclusion of $ \Delta$-isobar excitation. Short-distance physics is encoded in a few contact terms, the only parameters of the approach, fixed to reproduce selected properties of normal nuclear matter at zero temperature. The predicted nuclear thermodynamics  is then entirely determined by in-medium pion-nucleon dynamics with no additional free parameters. Correlations in the nuclear medium include two- and three-body forces.

Within this framework the nuclear equation of state  for different temperatures is calculated for a series of proton fractions $ x_p = Z/A $ in the full range $ 0 \leq x_p \leq 0.5 $ between pure neutron matter  and symmetric nuclear matter. The evolution of the first-order liquid-gas phase transition is systematically investigated. The critical temperature $ T_c \simeq 15.1 $ MeV at $ x_p = 0.5 $ falls as $ x_p $ decreases, and reaches $ T_c = 0 $ at $ x_p  \simeq 0.053 $, the point at which the liquid-gas transition disappears. The free energy is well approximated at low temperature by the usual quadratic expansion in the asymmetry parameter $ \delta = 1-2\,x_p $, even for large $ \delta $.

Part of the purpose of the present work is to set realistic constraints for ongoing discussions of the QCD phase diagram in the nuclear terrain of moderate baryon densities (up to about twice the density of normal nuclear matter) and temperatures up to about 50 MeV.  For example, nuclear chiral thermodynamics with its explicit treatment of pionic degrees of freedom in the nuclear medium is the appropriate framework that permits to calculate the density and temperature dependence of the chiral (quark) condensate beyond the Fermi gas limit \cite{kaiser7,kaiser8,kaiser9}, while respecting at the same time important nuclear physics constraints.

\section*{Acknowledgments}

We thank Jeremy W. Holt for many useful discussions.

\appendix 

\section{Contributions to the free energy density: \\ isospin-symmetric nuclear matter} \label{appA}

In this appendix we summarize the diagrammatic contributions to the free energy density by reporting the explicit expressions of the kernels for the convolution integrals Eq.~\eqref{convolution}. For further details and explanations see Refs.~\cite{kaiser2,kaiser3}.

\subsection*{One-body kernel}

\begin{equation}\label{one-body}
	\kn_1(p) = \tilde{\mu} - \frac{p^2}{3M_N} - \frac{p^4}{8M_N^3} \ ,
\end{equation}
where $ \tilde{\mu} $ is the effective one-body chemical potential and $ M_N = 939 $ MeV is the free nucleon mass.

\subsubsection*{Two-body kernels}

\begin{itemize}

\item Contact terms:
\begin{equation} \label{ct}
\kn_2^{(ct)}(\po,\pt) = 24  \pi^2 B_3 \frac{\po\,\pt}{M_N^2} + 20 \pi^2 B_5 \frac{\po\,\pt}{M_N^4} (\po^2+\pt^2) \ ,
\end{equation}
with $ B_3 = -7.99 $ and $ B_5 = 0 $.

\item $ 1\pi$-exchange Fock-diagram: 
\begin{multline}\label{1pifock}
\kn_2^{(1\pi)}(\po,\pt) = \frac{3\, \ga^2}{16 \fpi^2} \left \{ 8 \, \po \,\pt - 2 \, \mpi^2 \ln{\frac{\mpi^2 + (\otp)^2}{\mpi^2 + (\otm)^2}} \right. \\
\left. + \frac{1}{M_N^2} \left [ -4\,\po\,\pt (\po^2 + \pt^2) + \mpi^2 (\po^2 + \pt^2) \ln{\frac{\mpi^2 + (\po+\pt)^2}{\mpi^2 + (\po -\pt)^2}}	\right . \right. \\
\left. \left. - \frac{2\,\mpi^2\,\po \,\pt (\po^2 - \pt^2)^2}{[ \mpi^2 + (\otp)^2 ] [ \mpi^2 + (\otm)^2 ]}    \right ] \right \} \ .
\end{multline}

\item Iterated $ 1\pi$-exchange Hartree-diagram:
\begin{multline}\label{it1pihartree}
 \kn_2^{(\text{it} H)}(\po,\pt) = \frac{3\,\ga^4 M_N \mpi^2}{8 \pi \fpi^4} \left \{ (\po + \pt) \arctan{\frac{\po+\pt}{\mpi}} \right.  \\
 \left. + (\tom) \arctan{\frac{\otm}{\mpi}} - \frac{5}{8} \mpi \ln{\frac{\mpi^2+(\otp)^2}{\mpi^2+(\otm)^2}} \right \} \ .
\end{multline}

\item Iterated $ 1\pi$-exchange Fock-diagram:
\begin{multline}\label{it1pifock}
\kn_2^{(\text{it} F)}(\po,\pt) = \frac{3\,\ga^4 M_N \mpi}{32 \pi \fpi^4} \left \{ 2\,\po\,\pt + \mpi^2 \int\limits_{|\otm|/2\mpi}^{(\otp)/2\mpi} \frac{\mbox{d}x}{1+2x^2} \left [ (1+8x^2+8x^4) \arctan{x} \right . \right . \\
\left . \left . - (1+4x^2) \arctan{2x} \right ] \right\} \ .
\end{multline}

\item Irreducible $ 2\pi$-exchange Fock-diagram:
\begin{multline} \label{im}
\kn_2^{(\text{irr} F)}(\po,\pt) = \frac{1}{\pi} \int\limits_{2\mpi}^{\infty} \mbox{d}\mu \,\mbox{Im} (V_C + 3 W_C + 2\mu^2V_T + 6\mu^2 W_T) \\
\times \left \{ \mu \ln{\frac{\mu^2+(\otp)^2}{\mu^2+(\otm)^2}} - \frac{4\,\po\,\pt}{\mu} +  \frac{4\,\po\,\pt}{\mu^3} (\po^2+\pt^2) \right \} \ ,
\end{multline}
where $ \mbox{Im}V_C, \mbox{Im}W_C, \mbox{Im}V_T, \mbox{Im}W_T $ are the spectral functions of the isoscalar and isovector central and tensor $ NN$-amplitudes (see section 3 in Ref.~\cite{kaiser5}).

\end{itemize}

\subsubsection*{Three-body kernels}

\begin{itemize}

\item Iterated $ 1\pi$-exchange Hartree-diagram:
\begin{equation}\label{it1piHartree_3}
\kn_3^{(\text{it} H)}(\po,\pt,\pth) = \frac{3\,\ga^4 M_N}{4\fpi^4} \int\limits_{|\otm|}^{\otp} \mbox{d}q \frac{q^4}{(\mpi^2+q^2)^2} \ln{\frac{|\po^2-\pt^2+q^2+2\, \pth q|}{|\po^2-\pt^2+q^2-2\, \pth q|}} \ . 
\end{equation}

\item Iterated $ 1\pi$-exchange Fock-diagram:
\begin{multline}\label{it1pifock_3}
\kn_3^{(\text{it} F)}(\po,\pt,\pth) = \frac{3\,\ga^4 M_N}{16\fpi^4} \left\{ \frac{1}{8 \, \pth^3} \left[ 4\,\po\pth + \left( \pth^2-\po^2-\mpi^2 \right) \ln{\frac{\mpi^2+(\othp)^2}{\mpi^2+(\othm)^2}} \right] \right. \\
\times \left[ 4\,\pt\pth + \left(\pth^2-\pt^2-\mpi^2 \right) \ln{\frac{\mpi^2+(\tthp)^2}{\mpi^2+(\tthm)^2}} \right] \\
\left. + \int\limits_{|\pt-\pth|}^{\pt+\pth} \mbox{d}q\, \frac{q^2}{\mpi^2+q^2} \left[ \ln{\frac{|\po+h|}{|\pt-h|}} + \frac{\mpi^2}{R} \ln{\frac{|\po R + (\po^2-\pth^2-\mpi^2)h|}{|\po R + (\pth^2 +\mpi^2-\po^2)h|}} \right] \right\} \ ,
\end{multline}
\begin{alignat}{2}
h & = \frac{1}{2q}\left( \pt^2 - \pth^2 -q^2 \right) \ , & \qquad  R & = \sqrt{\left( \mpi^2+\po^2-\pth^2 \right)^2 + 4\mpi^2 \left(\pth^2-h^2 \right)} \ .
\end{alignat}

\item $ 2\pi$-exchange three body Hartree-diagram with single $ \Delta$-isobar excitation:
\begin{multline} \label{zeta}
\kn_3^{(\Delta\, H)}(\po,\pt,\pth) = \frac{3\,\ga^4 \pth}{\Delta \fpi^4} \left\{ 2\,\po\pt \left( 1+\zeta \right) + \frac{2\mpi^4 \,\po\pt}{\left[ \mpi^2 +(\otp)^2 \right] \left[ \mpi^2 + (\otm)^2 \right]}  \right . \\
\left. - \mpi^2 \ln{\frac{\mpi^2+(\otp)^2}{\mpi^2+(\otm)^2}} \right\} \ ,
\end{multline}
including the contact term with $ \zeta = - 3/4 $.

\item $ 2\pi$-exchange three body  Fock-diagram with single $ \Delta$-isobar excitation:
\begin{gather}
\kn_3^{(\Delta\, F)}(p_1,p_2,p_3) = - \frac{\ga^4}{4\Delta \fpi^4 \,\pth} \left[ 2\,X(\po,\pth) X(\pt,\pth) + Y(\po,\pth) Y(\pt,\pth) \right] \ ,\label{2pifock_3} \\
X(\po,\pth) = 2 \,\po\pth - \frac{\mpi^2}{2} \ln{\frac{\mpi^2+(\othp)^2}{\mpi^2+(\othm)^2}} \ , \label{X}\\
Y(\po,\pth) = \frac{\po}{4\,\pth} \left( 5\,\pth^2 - 3\mpi^2 - 3\,\po^2 \right) 
+ \frac{3 \,(\po^2-\pth^2+\mpi^2)^2 + 4\mpi^2 \,\pth^2}{16 \,\pth^2} \ln{\frac{\mpi^2+(\othp)^2}{\mpi^2+(\othm)^2}} \ . \label{Y}
\end{gather}

\end{itemize}

\subsubsection*{Anomalous contribution}

\begin{multline}\label{anomalous}
\rho \mathcal{\bar{A}}(\rho, T) = - \frac{\left[ \Omega'_{1\pi}(\rho, T) \right]^2}{2 \Omega''_0(\rho, T)} + \frac{9\,\ga^4}{8\fpi^4 T} \int\limits_0^\infty \mbox{d}\po  \int\limits_0^\infty \mbox{d}\pt  \int\limits_0^\infty \mbox{d}\pth \,d(\po) \,d(\pt) \left[ 2\pi^2 d(\pt) - \pt \right] d(\pth) \\
\times \left[ \po - \frac{\mpi^2}{4\,\pt} \ln{\frac{\mpi^2+(\otp)^2}{\mpi^2+(\otm)^2}} \right] \left[ \pth - \frac{\mpi^2}{4\,\pt} \ln{\frac{\mpi^2+(\pth+\pt)^2}{\mpi^2+(\pth-\pt)^2}} \right] \ ,
 \end{multline}
with 
\begin{gather}
\Omega'_{1\pi}(\rho, T) = \frac{3\,\ga^2 M_N}{2\fpi^2} \int\limits_0^\infty \mbox{d}\po \int\limits_0^\infty \mbox{d}\pt \, d(\po) \frac{d(\pt)}{\pt} \left[ \frac{(\po+\pt)^3}{\mpi^2+(\otp)^2} + \frac{(\otm)^3}{\mpi^2+(\otm)^2} \right] \ , \label{om1pi}\\
\Omega''_0(\rho, T) = - 4 M_N \int\limits_0^\infty \mbox{d}p\, \frac{d(p)}{p} \ .
\end{gather}

\section{Contributions to the free energy density:\\ isospin-asymmetric nuclear matter} \label{appB}

In isospin-asymmetric nuclear matter the Fermi momenta and the ``one-body'' chemical potentials of protons and neutrons are different. As a consequence, Eq.~\eqref{convolution} has to be modified, by taking into account that
\begin{itemize}
\item for each closed diagram we sum over all possible combinations of protons and neutrons, each combination multiplied by its own isospin factor;
\item the densities of states $ d_{p}(q) $ and $ d_n(q) $ are different due to different $ \tilde{\mu}_p $ and $ \tilde{\mu}_n $. 
Note that the momenta to be integrated over are now denoted by the symbol $ q $ instead of $ p $.
\end{itemize}

\subsubsection*{One-body kernel}

The integration over the proton and neutron distributions have to be performed separately, resulting in the following replacement in Eqs.~\eqref{convolution} and \eqref{one-body}:
\begin{equation} \label{k1}
4 \int\limits_0^\infty \mbox{d}q \, q\, \kn_1(q) d(q) \longrightarrow 2 \int\limits_0^\infty \text{d}q \, q \left[ \kn_1^{(p)}(q) d_p(q) + \kn_1^{(n)}(q) d_n(q) \right] \ ,
\end{equation}   
where $ \displaystyle \kn_1^{(p,n)}(q) = \tilde{\mu}_{p,n} - \frac{q^2}{3M_N} -\frac{q^4}{8M_N^3} $.

\subsubsection*{Two-body kernels}

\begin{itemize}

\item  $ 1\pi$-exchange Fock-diagram. \\
Using $ \kn_2^{(1\pi)} $ of Eq.~\eqref{1pifock} the product $ d(q_1) d(q_2 ) $ for isospin-symmetric nuclear matter is replaced by:
\begin{equation}
 d(q_1) d(q_2) \longrightarrow \frac{1}{6}  \left[ d_p(q_1) d_p(q_2) + d_n(q_1) d_n(q_2) + 4\, d_p(q_1) d_n(q_2) \right] \ .
\end{equation}

\item Iterated $ 1\pi$-exchange Hartree-diagram. \\
Using $ \kn_2^{(\text{it} H)} $ from Eq.~\eqref{it1pihartree}, replace in the corresponding integral: 
\begin{equation}
d(q_1) \, d(q_2)  \longrightarrow \frac{1}{12}  \left[ d_p(q_1) \,d_p(q_2) + d_n(q_1)\,d_n(q_2) + 10\, d_p(q_1)\,d_n(q_2) \right] \ .
\end{equation}

\item Iterated $ 1\pi$-exchange Fock-diagram.\\
Using $ \kn_2^{(\text{it} F)} $ from Eq.~\eqref{it1pifock}, replace in the corresponding integral:
\begin{equation}
d(q_1)\,d(q_2) \longrightarrow  \frac{1}{6} \left[ 8\, d_p(q_1)\,d_n(q_2)- d_p(q_1)\,d_p(q_2) - d_n(q_2)\,d_n(q_2)  \right] \ .
\end{equation}

\item Irreducible $ 2\pi$-exchange.\\
For isoscalar amplitude $ V_{C,T} $ in Eq.~\eqref{im}, replace:
\begin{equation}
d(q_1)\,d(q_2) \longrightarrow \frac{1}{2} \left[ d_p(q_1)\,d_p(q_2) + d_n(q_1)\,d_n(q_2) \right]\ .
\end{equation}
For isovector amplitude $ W_{C,T} $ in Eq.~\eqref{im}, replace:
\begin{equation}
 d(q_1)\,d(q_2) \longrightarrow \frac{1}{6} \left[ d_p(q_1)\,d_p(q_2) + d_n(q_1)\,d_n(q_2) + 4 \, d_p(q_1) d_n(q_2) \right] \ .
\end{equation}

\item Contact terms.\\
Expressions for symmetric nuclear matter using $ \kn_2^{(ct)} $ of Eq.~\eqref{ct} are replaced as follows:
\begin{eqnarray}
B_3 \,d(q_1)\,d(q_2) \longrightarrow (B_3-B_{n,3} ) \,d_p(q_1)\, d_n(q_2) + \frac{1}{2} B_{n,3} \Big[ d_p(q_1)\,d_p(q_2) + d_n(q_1)\,d_n(q_2) \Big] \ ,\\
B_5 \,d(q_1)\,d(q_2) \longrightarrow (B_5-B_{n,5} )\, d_p(q_1)\, d_n(q_2) + \frac{1}{2} B_{n,5} \Big[ d_p(q_1)\,d_p(q_2) + d_n(q_1)\,d_n(q_2) \Big]\ ,
\end{eqnarray}
with $ B_{n,3} = -0.95 $, $ B_{n,5} = -3.58 $.

\end{itemize}

\subsubsection*{Three-body kernels}

\begin{itemize}

\item Iterated $ 1\pi$-exchange Hartree-diagram.\\
Using $  \kn_3^{(\text{it} H)}  $ from Eq.~\eqref{it1piHartree_3}, replace in the corresponding integral:
\begin{multline}
d(q_1)\,d(q_2)\,d(q_3) \longrightarrow \frac{1}{12} \Big[ d_p(q_1)\,d_p(q_2)\,d_p(q_3) + d_n(q_1)\,d_n(q_2)\,d_n(q_3) +d_p(q_1)\,d_p(q_2)\,d_n(q_3) \\ + d_n(q_1)\,d_n(q_2)\,d_p(q_3) + 4\,d_n(q_1)\,d_p(q_2)\,d_n(q_3) + 4\,d_p(q_1)\,d_n(q_2)\,d_p(q_3) \Big] \ .
\end{multline}

\item Iterated $ 1\pi$-exchange Fock-diagram. \\
Using $  \kn_3^{(\text{it} F)}  $ from Eq.~\eqref{it1pifock_3} , replace in the corresponding integral:
\begin{multline}
d(q_1)\,d(q_2)\,d(q_3) \longrightarrow \frac{1}{6} \Big[ 2\,d_p(q_1)\,d_n(q_2)\,d_p(q_3) + 2\, d_p(q_1)\,d_n(q_2)\,d_n(q_3)- d_p(q_1)\,d_p(q_2)\,d_p(q_3) \\+ 2\,d_n(q_1)\,d_p(q_2)\,d_p(q_3) + 2\,d_n(q_1)\,d_p(q_2)\,d_n(q_3) - d_n(q_1)\,d_n(q_2)\,d_n(q_3) \Big]\ .
\end{multline}

\item Irreducible $ 2\pi$-exchange Hartree-diagram with single $ \Delta$-isobar excitation.\\
Using $  \kn_3^{(\Delta\,H)}  $ from Eq.~\eqref{zeta}, replace in the corresponding integral:
\begin{multline}
d(q_1)\,d(q_2)\,d(q_3) \longrightarrow\frac{1}{12} \Big[d_p(q_3) + d_n(q_3)\Big] \Big[ d_p(q_1)\,d_p(q_2)+d_n(q_1)\,d_n(q_2) +4\,d_p(q_1)\,d_n(q_2)\Big] \ ,
\end{multline}
The three-body contact term proportional to $ \zeta $ is excluded from this formula,  because it transforms differently:
\begin{equation}
\zeta  \,d(q_1)\,d(q_2)\,d(q_3) \longrightarrow  \zeta \,\frac{1}{2}  d_p(q_1)\,d_n(q_2) [d_p(q_3)+d_n(q_3) ] \ .
\end{equation}

\item Irreducible $ 2\pi$-exchange Fock-diagram with single $ \Delta$-isobar excitation.\\
Expressions for symmetric nuclear matter using $ \kn_3^{(\Delta\,F)} $ of Eq.~\eqref{2pifock_3} are replaced as follows:
\begin{multline}
2\,X(q_1,q_3)X(q_2,q_3) \,d(q_1)\,d(q_2)\,d(q_3) \longrightarrow   \frac{1}{6}   X(q_1,q_3) X(q_2,q_3) \,\times \\
\Big[ d_p(q_1)\,d_p(q_2)\,d_p(q_3) + d_p(q_1)\,d_n(q_2)\,d_p(q_3) + d_n(q_1)\,d_p(q_2)\,d_p(q_3) + 3\,d_n(q_1)\,d_n(q_2)\,d_p(q_3) \\+ d_n(q_1)\,d_n(q_2)\,d_n(q_3) + d_n(q_1)\,d_p(q_2)\,d_n(q_3) + d_p(q_1)\,d_n(q_2)\,d_n(q_3) + 3\,d_p(q_1)\,d_p(q_2)\,d_n(q_3) \Big] \ , \nonumber
\end{multline}
\vspace{-0.4cm}
\begin{multline}
 Y(q_1,q_3)Y(q_2,q_3)  \,d(q_1)\,d(q_2)\,d(q_3) \longrightarrow \ \frac{1}{6}  Y(q_1,q_3) Y(q_2,q_3) \,\times \\ \Big[  2\,d_p(q_1)\,d_p(q_2)\,d_p(q_3)-d_p(q_1)\,d_n(q_2)\,d_p(q_3) - d_n(q_1)\,d_p(q_2)\,d_p(q_3) 
+3\,d_n(q_1)\,d_n(q_2)\,d_p(q_3)\\ +2\,d_n(q_1)\,d_n(q_2)\,d_n(q_3) - d_n(q_1)\,d_p(q_2)\,d_n(q_3)-d_p(q_1)\,d_n(q_2)\,d_n(q_3) +3\,d_p(q_1)\,d_p(q_2)\,d_n(q_3)\Big]  \ ,
\end{multline}
where $ X(q_1,q_3) $ and $ Y(q_1,q_3) $ are given in Eq.~\eqref{X} and \eqref{Y}.
\end{itemize}

\subsubsection*{Anomalous term}

The anomalous term is given by the sum of two terms  transforming in a different way.
\begin{itemize}

\item Replace in the triple integral in Eq.~\eqref{anomalous}:
\begin{multline}
d(q_1)\,d(q_2) \left[ 2\pi^2 d(q_2) - q_2 \right] d(q_3) \longrightarrow\\ \frac{1}{18} \bigg\{ d_p(q_1)\,d_p(q_3) \big\{ d_p(q_2) [ 2\pi^2d_p(q_2) -q_2 ] + 4\,d_n(q_2)[2\pi^2d_n(q_2)-q_2] \big\} \\
+ d_n(q_1)\,d_n(q_3) \big\{ d_n(q_2) [ 2\pi^2d_n(q_2) -q_2 ] + 4\,d_p(q_2)[2\pi^2d_p(q_2)-q_2] \big\} \\
+ 4\, d_p(q_1)\,d_n(q_3) 
 \big\{ d_p(q_2) [ 2\pi^2d_p(q_2) -q_2 ] + d_n(q_2)[2\pi^2d_n(q_2)-q_2] \big\} \bigg\} \ .
\end{multline}

\item The subtraction term in Eq.~\eqref{anomalous} transforms as:
\begin{equation}
- \frac{\left[ \Omega'_{1\pi}(\rho, T) \right]^2}{2\, \Omega''_0(\rho, T)} \longrightarrow
- \frac{\left[ \Omega'_{1\pi p}(\rho_p,\rho_n, T) \right]^2}{2\, \Omega''_{0p}(\rho_p,T)}
- \frac{\left[ \Omega'_{1\pi n}(\rho_p,\rho_n, T) \right]^2}{2\, \Omega''_{0n}(\rho_n, T)}
\end{equation}
with $ \rho = \rho_p+\rho_n $ and
\begin{align}
\Omega''_{0p,n}(\rho_{p,n}, T)  = - 2 M_N \int\limits_0^\infty \mbox{d}q\, \frac{d_{p,n}(q)}{q}\ . 
\end{align}
Using $ \Omega'_{1\pi}(\rho, T) $ defined in Eq.~\eqref{om1pi}, $ \Omega'_{1\pi p} $ is obtained replacing in the integral:
\begin{align}
d(q_1)\,d(q_2)  \longrightarrow  \frac{1}{6} \big[ d_p(q_1)\,d_p(q_2)+2\,d_n(q_1)\,d_p(q_2) \Big] \ .
\end{align}ÇÇÇ
Using $ \Omega'_{1\pi}(\rho, T) $ defined in Eq.~\eqref{om1pi}, $ \Omega'_{1\pi n} $ is obtained replacing in the integral:
\begin{equation}
d(q_1)\,d(q_2)   \longrightarrow  \frac{1}{6} \big[ d_n(q_1)\,d_n(q_2)+2\,d_p(q_1)\,d_n(q_2) \Big] \ .
\end{equation}

\end{itemize}

\end{document}